\documentclass[aps,prd,twocolumn,showpacs,preprintnumbers,superscriptaddress,nofootinbib,floatfix]{revtex4-1}

\usepackage{amsmath}
\usepackage{bm}
\usepackage{graphicx}
\usepackage{epstopdf}
\usepackage{hyperref}
\usepackage{array}

\usepackage{soul}
\usepackage{color}

\enlargethispage{\baselineskip}

\hypersetup{
     colorlinks   = true,
     citecolor    = blue
}

\newcolumntype{C}[1]{>{\centering\let\newline\\\arraybackslash\hspace{0pt}}m{#1}}
\newcommand{\fc}{f_\chi}
\newcommand{\SI}{\sigma^{\rm SI}_p}
\newcommand{\SDp}{\sigma^{\rm SD}_p}
\newcommand{\SDn}{\sigma^{\rm SD}_n}
\newcommand{\tf}{\tilde{f}}
\newcommand{\Ve}{V_{\rm eff}}
\newcommand{\Nni}{N^X_{\nu_{\ell'}}}
\newcommand{\Pni}{\Phi^{\rm DM}_{\nu_{\ell}}}
\newcommand{\mB}{\mathcal{B}}

\newcommand{\mP}{\mathcal{P}}
\newcommand{\Em}{E_{\rm min}}
\newcommand{\EM}{E_{\rm max}}
\newcommand{\br}{{\bf r}}
\newcommand{\bx}{{\bf x}}
\newcommand{\bu}{{\bf u}}

\newcommand{\ve}{v_{\rm esc}}
\newcommand{\n}{{\bf \nabla}}
\newcommand{\mF}{\mathcal{F}}

\newcommand{\FSI}{F_{\rm SI}}
\newcommand{\FSD}{F_{\rm SD}}

\newcommand{\sann}{\langle\,\sigma\,v\rangle_{\rm ann}}
\newcommand{\tann}{\tau_{\rm ann}}

\newcommand{\Sp}{\langle S_p^i\rangle}
\newcommand{\Sn}{\langle S_n^i\rangle}

\begin{document}

\newcommand{\FIRSTAFF}{\affiliation{The Oskar Klein Centre for Cosmoparticle Physics,
	Department of Physics,
	Stockholm University,
	AlbaNova,
	10691 Stockholm,
	Sweden}}
\newcommand{\SECONDAFF}{\affiliation{Nordita,
	KTH Royal Institute of Technology and Stockholm University
	Roslagstullsbacken 23,
	10691 Stockholm,
	Sweden}}
\newcommand{\THIRDAFF}{\affiliation{University of Helsinki,
	P.O. Box 64, FI-00014
	Finland.}}
\newcommand{\FOURTHAFF}{\affiliation{Departement of Physics,
		University of Michigan,
		Ann Arbor, MI 48109, USA}}

\title{A tale of dark matter capture, sub-dominant WIMPs, and neutrino observatories}

\author{Sebastian Baum}
\email[Electronic address: ]{sbaum@fysik.su.se}
\FIRSTAFF
\SECONDAFF

\author{Luca Visinelli}
\email[Electronic address: ]{luca.visinelli@fysik.su.se}
\FIRSTAFF
\SECONDAFF
\THIRDAFF

\author{Katherine Freese}
\email[Electronic address: ]{ktfreese@umich.edu}
\FIRSTAFF
\SECONDAFF
\FOURTHAFF

\author{Patrick Stengel}
\email[Electronic address: ]{pstengel@umich.edu}
\FOURTHAFF

\preprint{NORDITA-2016-130}
\preprint{MCTP-16-32}

\date{\today}

\begin{abstract}
Weakly Interacting Massive Particles (WIMPs), which are among the best motivated dark matter (DM) candidates, could make up all or only a fraction of the total DM budget. We consider a scenario in which WIMPs are a sub-dominant DM component; such a scenario would affect both current direct and indirect bounds on the WIMP-nucleon scattering cross section. In this paper we focus on indirect searches for the neutrino flux produced by annihilation of sub-dominant WIMPs captured by the Sun or the Earth via either spin-dependent or spin-independent scattering.  We derive the annihilation rate and the expected neutrino flux at neutrino observatories.  In our computation, we include an updated chemical composition of the Earth with respect to the previous literature, leading to an increase of the Earth's capture rate for spin-dependent scattering by a factor three. Results are compared with current bounds from Super-Kamiokande and IceCube. We discuss the scaling of bounds from both direct and indirect detection methods with the WIMP abundance.
\end{abstract}

\maketitle

\section{Introduction}

Weakly interacting massive particles (WIMPs) are among the best motivated candidates to explain the observed dark matter (DM). WIMPs naturally occur in extensions of the Standard Model, e.g. the lightest neutralino in supersymmetric extensions of the Standard Model, the lightest Kaluza-Klein photon in universal extra-dimension theories, and the heavy photon in Little Higgs models. WIMPs can be produced in the early universe with relic density matching the observed DM energy density, e.g. via the freeze-out mechanism~\cite{lee1977, hut1977, sato1977}. Current searches involve both direct and indirect detection, as well as accelerator searches.
For reviews of approaches to WIMP detection, see Refs.~\cite{primack1988, lewin1996, jungman1996, bertone2005, freese2013}.

Since this plethora of searches has not yet yielded conclusive evidence for the existence of WIMPs, recent years have seen the development of model-independent techniques to analyze those null-results. Namely, the non-relativistic effective field theory (EFT) framework~\cite{fitzpatrick2013} has been developed for direct detection and simplified models are employed in recent analyses of bounds from the Large Hadron Collider~(cf. \cite{Abercrombie:2015wmb} and references therein).

If WIMPs exist, they may accumulate~\cite{press1985} in the Earth~\cite{freese1986, krauss1986, gould1987a, gould1989} and in the Sun~\cite{silk1985, gaisser1986b, srednicki1987, griest1987, gould1987b} via down-scattering off the body's material. 
The first paper to point out that annihilation in the Sun can lead to a detectable neutrino signal was by Silk, Olive, and Srednicki \cite{silk1985}; the first papers to point out that annihilation in the Earth can lead to a detectable neutrino signal in the Earth were
by Freese \cite{freese1986} and Krauss and Wilczek \cite{krauss1986}.
As shown by these authors, the captured WIMP population could then annihilate and give rise to a flux of energetic neutrinos, which may be detectable at neutrino observatories such as the Super-Kamiokande (Super-K)~\cite{desai2004, tanaka2011a, tanaka2011b, choi2015a, choi2015b}, IceCube~\cite{abbasi2011, aartsen2016a, aartsen2016b}, ANTARES~\cite{aslanides1999, zornoza2016}, and AMANDA~\cite{andres2000} facilities, or in the proposed KM3NeT neutrino telescope~\cite{adrian2016}. DM capture and annihilation in the Sun and Earth has recently also been used to constrain inelastic and self-interacting DM models~\cite{menon2009, nussinov2009, andreas2009, schuster2010, taoso2010, fan2014, Kouvaris2015, chen2015, blennow2015, dev2016, feng2016, murase2016, catena2016b}, supersymmetric models~\cite{barger2007}, and DM models with a boosted annihilation cross section~\cite{delaunay2009}.

Both direct detection experiments and the annihilation rate of captured WIMPs are sensitive to the local WIMP energy density $\rho^{\rm loc}_{\chi}$. Models where WIMPs constitute only a fraction of the total DM budget~\cite{bottino2000, bottino2001, duda2002, duda2003, gelmini2005, profumo2009, blum2014} have local WIMP densities different from the measured local DM density. This must be taken into account when considering bounds from direct and indirect detection. 

We consider two scenarios: i) WIMPs comprising all of the observed DM, or ii) a sub-dominant fraction of DM only. For both scenarios, we derive the annihilation rate and the induced neutrino flux from capture and annihilation in the Sun or the Earth, and compare the induced neutrino flux with current bounds from neutrino observatories. We introduce minimal assumptions on the nature of the WIMP particle. In particular, we do not assume that the scattering and the annihilation cross sections are related by a crossing symmetry but we treat them as independent parameters. The annihilation cross-section is fixed by demanding that the DM fraction in WIMPs is obtained through a thermal freeze-out mechanism. We update the composition of the Earth with respect to current literature to include various isotopes that are important for spin-dependent capture.

This paper is organized as follows. In Sec.~\ref{WIMP capture and annihilation by a massive body} we review the expected neutrino flux from WIMP annihilation in the Sun and in the Earth, we update the chemical composition of the Earth to obtain new result on the SD WIMP capture rate, and we compute the portion of the parameter space for which the capture process is in equilibrium in the Sun and the Earth for both spin-independent (SI) and spin-dependent (SD) interactions, comparing results with current bounds from direct detection. Sec.~\ref{Subdominant WIMP DM model} is devoted to analyzing the effect of a sub-dominant WIMP fraction of the DM on the annihilation rate. In Sec.~\ref{Muon flux at Super-K and IceCube}, we use the updated values of the muon flux from muon neutrinos at the detector site to constrain the SI and SD cross sections as a function of the WIMP mass. In particular, in Sec. \ref{sec:scaling} we give a detailed discussion of the scaling behavior of signals and bounds from DM capture and annihilation for WIMPs comprising a fraction of the DM only, and in section \ref{Effect of updated composition of the Earth} we discuss the effect of the updated chemical composition of the Earth on the constraints. 

\section{WIMP capture and annihilation by a massive body} \label{WIMP capture and annihilation by a massive body}

The capture rate of WIMPs by a massive body has first been estimated in Ref.~\cite{press1985}, and has immediately been applied to capture by the Earth in Refs.~\cite{freese1986, krauss1986, gould1987a}, and by the Sun in Refs.~\cite{silk1985, gaisser1986b, srednicki1987, griest1987, gould1987b}. A massive body, like the Earth or the Sun, builds up a population of WIMPs at a rate $C$ by capturing them via scattering off the body's nuclei. A particle is said to be captured if its velocity is smaller than the escape velocity $\ve$ of the capturing body.

WIMPs captured in the massive body can annihilate at a rate $\Gamma_A$, which is given by the number density profile of capture WIMPs $n(\br, t)$ and the velocity-averaged annihilation cross section $\sann$ as
\begin{equation} \label{annihilation_rate0}
\Gamma_A = \sann\,\int \,n^2(\br, t)\, d^3\br.
\end{equation}
 
Besides via self-annihilation, the population of WIMPs captured in the body may also be depleted by evaporation at the rate $C_E$, if captured WIMPs regain enough energy to escape the gravitational potential of the body via hard scattering with nuclei ~\cite{nauenberg1987, griest1987, gould1987c, gould1990a, gould1990b}. 

The total number of WIMPs $N(t)$ captured by a massive body after time $t$ is given by the solution to the differential equation
\begin{equation} \label{diff_eq_numbercapturewimps}
\frac{dN}{dt} = -C_A\,N^2 - C_E\,N + C,
\end{equation}
where the constant $C_A$ is related to $\Gamma_A$ and to the number of captured WIMPs by
\begin{equation} \label{annihilation_rate}
\Gamma_A = \frac{C_A}{2}\,N^2(t).
\end{equation}

Eq. \eqref{diff_eq_numbercapturewimps} assumes that WIMPs, once they are captured, thermalize on time scales much shorter than the age of the solar system, which allows to separate the $t$- and $\br$-dependence of the number density profile as~\cite{peter2009a, peter2009b}
\begin{equation} \label{number_density}
n(\br, t) = N(t)\,\tilde{n}(\br) \equiv N(t)\,\frac{e^{-m_\chi\,\Phi(r)/T}}{\int e^{-m_\chi\,\Phi(r)/T}\,d^3\br},
\end{equation}
where $\tilde{n}(\br) = n(\br,t)/N(t)$ is the normalized number density profile, which is determined by the gravitational potential of the capturing body $\Phi(r)$ and the body's temperature profile $T(r)$~\cite{freese1986, gould1987a, gould1987b, edsjo1995}. In the case of capture in the Sun, it has been shown that thermalization time scales are shorter than capture time scales if the SI (SD) WIMP-proton scattering cross section satisfies $\SI \gtrsim 10^{-48}\,$cm$^2$ ($\SDp \gtrsim 10^{-51}\,$cm$^2$) for WIMP masses $m_\chi \approx 100\,$GeV~\cite{peter2009a, peter2009b}. In this work, we make the assumption that thermalization proceeds much faster than capture for the entire WIMP parameter space considered.

From Eq.~\eqref{diff_eq_numbercapturewimps}, we obtain the time evolution of the number of WIMPs as
\begin{equation} \label{number_solution}
N(t) = \sqrt{\frac{C}{C_A}}\,\frac{\tanh\left(\frac{\alpha\,t}{\tann}\right)}{\alpha+\sqrt{\alpha^2-1}\,\tanh\left(\frac{\alpha\,t}{\tann}\right)},
\end{equation}
where $\tann \equiv 1/\sqrt{C\,C_A}$ is the time scale after which the capture and annihilation processes reach equilibrium, and $\alpha \equiv \sqrt{1+(C_E\,\tann/2)^2}$. It has been shown, that for WIMP masses $m_\chi \gtrsim 5\,$GeV considered in this work, evaporation can be neglected for both the Sun and the Earth~\cite{nauenberg1987, gould1987c, gould1990a, gould1990b, kappl2011}. In this case, $\alpha \to 1$ and Eq.~\eqref{number_solution} reduces to
\begin{equation} \label{number_solution1}
N(t) = \sqrt{\frac{C}{C_A}}\,\tanh\left(\frac{t}{\tann}\right),
\end{equation}
which is the expression we use in our numerical computation. Inserting Eq.~\eqref{number_solution1} into Eq.~\eqref{annihilation_rate} gives the present annihilation rate
\begin{equation} \label{annihilation_rate_time}
\Gamma_A = \frac{C}{2}\,\tanh^2\left(\frac{t_{\odot}}{\tann}\right),
\end{equation}
where $t_{\odot}$ is the age of the solar system, and the equilibrium time scale is given by
\begin{equation} \label{relaxation_time}
\tann = \sqrt{\frac{\Ve}{C\,\sann}},
\end{equation}
where the effective volume $\Ve$ is given in Appendix~\ref{annihilation_rate_rev}, Eq.~\eqref{effective_volume}.

WIMP annihilation leads to a differential flux of neutrinos of flavor $\ell = e, \mu, \tau$ as~\cite{edsjo1995, edsjo1997, bertone2005, strigari2009}
\begin{equation} \label{diff_neutrino_flux}
\frac{d\,\Pni}{dE_\nu} = \frac{\Gamma_A}{4\pi\,D^2}\,\sum_{\ell'} \,\mP_{\nu_{\ell'}\to\nu_\ell}(E_\nu, D)\,\sum_X\,\mB_\chi^X\,\frac{d\Nni}{dE_\nu}.
\end{equation}
Here, $\mB_\chi^X$ is the branching ratio for the DM annihilation channel $\chi \bar{\chi} \to X\bar{X}$, and $\mP_{\nu_{\ell'}\to\nu_\ell}(E_\nu, D)$ is the probability that a neutrino converts from the species $\ell'$ to the species $\ell$ along the distance $D$ between the source and the detector. $d\Nni/dE_\nu$ is the neutrino spectrum obtained from the decay chain of $X$.

\subsection{WIMP capture rate by a massive body} \label{Dark matter capture rate by a massive body}

We give a detailed review of the calculation of the capture rate $C$ in Appendix~\ref{Review of the WIMP capture rate}. We write the capture rate (cf. Eq. \eqref{capture_rate_general}) as
\begin{equation} \label{capture_rate_text}
C = K^s(m_\chi)\,\sigma^s_p\,\rho^{\rm loc}_\chi.
\end{equation}
Here, $\sigma^s_p$ is the WIMP-proton scattering cross section at zero momentum for either SI or SD scattering, $\rho^{\rm loc}_\chi$ is the local WIMP energy density, and the function $K^s(m_\chi)$ is defined in Eq.~\eqref{def_kappa}. We refer to Appendix~\ref{Review of the WIMP capture rate} for additional details on the notation used. In this work, we present results for isospin conserving WIMP-nucleon scattering $\sigma_p^s = \sigma_n^s$ for both SI and SD scattering and show our figures in the $\sigma_p^s - m_\chi$ plane. Changing the WIMP-neutron scattering cross-sections has little effect on WIMP capture and annihilation in the Sun, which is predominantly composed of hydrogen, i.e. protons. The Earth on the other hand is composed of a range of heavier elements, cf. Table~\ref{table_fraction_earth}. Assuming isospin violating WIMP-nucleon cross sections can have dramatic effects on the capture rate and hence on the corresponding bounds in the Earth. However, the isospin violation is model dependent and we remain agnostic about an underlying model for WIMPs. For this work, we choose to present results for the isospin conserving case only.

The function $K^s(m_\chi)$ strongly depends on the abundance and distribution of the chemical elements in the capturing body. For the Earth, we update the table found in the {\tt DarkSUSY} package, by including the abundances provided in Ref.~\cite{mcdonough2003} and summarized in Table~\ref{table_fraction_earth}, which is used in the recent literature in DM capture~\cite{catena2016}. We include all stable isotopes of the 14 most abundant elements in the Earth mantle and core; 35 isotopes in total. Of these nuclei, 13 give rise to spin-dependent scattering, namely $^1$H, $^{13}$C, $^{17}$O, $^{23}$Na, $^{25}$Mg, $^{27}$Al, $^{29}$Si, $^{31}$P, $^{43}$Ca, $^{53}$Cr, $^{55}$Mn, $^{57}$Fe, and $^{61}$Ni. The values of $\phi_i$ reported in Table~\ref{table_fraction_earth} for the Earth are taken from Ref.~\cite{gondolo2004}, except for carbon and hydrogen which are not provided, and which we compute via Eq.~\eqref{definition_phi}. For the Earth, we assume the mean isotopic distribution to be the same in both the mantle and the core, so that all isotopes of a given element have the same $\phi_i$. For the Sun, we use the abundances and the effective potential $\phi_i$ tabulated in {\tt DarkSUSY}~\cite{gondolo2004}, see Table~\ref{table_fraction_sun} (located in Appendix~\ref{Solar Capture}), which are based on the method outlined in Ref.~\cite{jungman1996} and the standard solar model~\cite{dalsgaard1996, bahcall2001, bahcall2005}.

\begin{table*}
\begin{center}
\begin{tabular}{|c|C{2cm}|C{2cm}|C{2cm}|c|}
\hline
Isotope & \multicolumn{3}{c|}{Mass Fraction} & Potential \\
$i$ & \multicolumn{3}{c|}{$x_i$ ($\%$)} & $\phi_i$ \\
\hline
& Mantle & Core & Total & \\
\hline
Fe & 6.26 & 85.5 & 32.0 & 1.59\\
O & 44.0 & 0.0 & 29.7 & 1.28\\
Si & 21.0 & 6.0 & 16.1 & 1.33\\
Mg & 22.8 & 0.0 & 15.4 & 1.28\\
Ni & 0.20 & 5.2 & 1.82 & 1.63\\
Ca & 2.53 & 0.0 & 1.71 & 1.28\\
Al & 2.35 & 0.0 & 1.59 & 1.28\\
S & 0.03 & 1.9 & 0.64 & 1.62\\
Cr & 0.26 & 0.9 & 0.47 & 1.50\\
Na & 0.27 & 0.0 & 0.18 & 1.30\\
P  & 0.009 & 0.2 & 0.07 & 1.63\\
Mn & 0.10 & 0.30 & 0.08 & 1.54\\
C & 0.01 & 0.20& 0.07 & 1.64\\
H & 0.01 & 0.06 & 0.03 & 1.35\\
\hline
\end{tabular}
\caption{Most abundant isotopes of the Earth mantle and core, together with their total mass fractions, as given in Ref.~\cite{mcdonough2003}. The potentials $\phi_i$ are from Ref.~\cite{gondolo2004}, except for carbon and hydrogen for which we have used Eq.~\eqref{definition_phi}.}
\label{table_fraction_earth}
\end{center}
\end{table*}

\subsection{Results for the capture rate}

\begin{figure*}
\includegraphics[width=\linewidth]{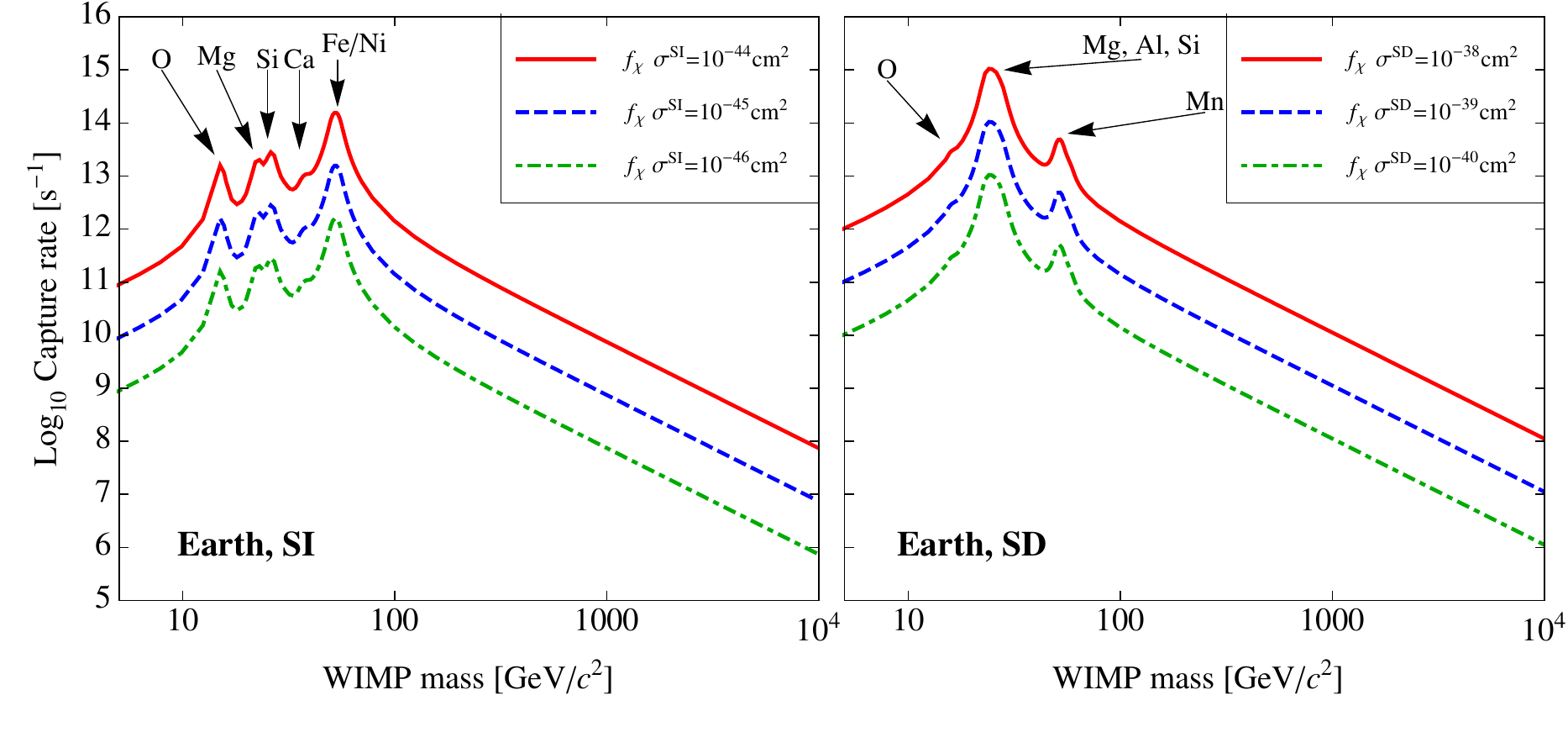}
\caption{The total capture rate (in s$^{-1}$) in the Earth, as a function of the DM mass~$m_\chi$, for different values of the WIMP-nucleon cross section. red: $\SI = 10^{-44}\,$cm$^2$ or $\SDp = 10^{-38}\,$cm$^2$; blue dashed: $\SI = 10^{-45}\,$cm$^2$ or $\SDp = 10^{-39}\,$cm$^2$; green dash-dotted: $\SI = 10^{-46}\,$cm$^2$ or $\SDp = 10^{-40}\,$cm$^2$. The parameter $\fc$ is introduced later in the text and gives the WIMP fraction of the total DM budget.}
\label{fig_rate_earth}
\end{figure*}

We show the capture rate for the Earth in Fig.~\ref{fig_rate_earth} and the Sun in Fig.~\ref{fig_rate_sun}, considering both SI (left) and SD scattering cross section (right). The results for the solar capture rate agree with previous findings in the literature;
the results for Earth capture use the updated elemental abundances in the Earth and are therefore improvements upon the previous
literature. The values for the proton-WIMP cross sections have been chosen to be compatible with the latest measurements by CDMS~\cite{agnese2014} and LUX~\cite{akerib2014} for SI and by PICO~\cite{amole2016} for SD, and we use a local DM energy density
\begin{equation} \label{DM_local}
\rho^{\rm loc}_{\rm DM} = 0.4\,{\rm GeV/cm^3}.
\end{equation}
One finds enhanced capture rates when the DM mass $m_\chi$ matches the mass of the nucleus it scatters off, see Ref.~\cite{gould1987b} for a discussion of this resonant enhancement. The width of this resonance is set by the ratio of the capturing body's escape velocity $\ve$ to the DM velocity dispersion $v_\sigma$. For the Sun, $\ve/v_\sigma \sim 2$ and one does not find pronounced features in the capture rate. For the Earth on the other hand, $\ve/v_\sigma \sim 0.04$ and one makes out a number of distinguished features: For SI scattering, the largest resonances are obtained for $^{16}$O, $^{28}$Si, and Fe/Ni, where the Fe/Ni peak is caused by overlapping contributions from $^{56}$Fe, $^{58}$Ni, and $^{60}$Ni. There are further less-pronounced peaks from scattering off $^{24}$Mg, $^{32}$S, and $^{40}$Ca. The SD capture rate for the Earth shows resonances for $^{55}$Mn and $^{25}$Mg, $^{27}$Al, and $^{29}$Si. The resonance peaks of the last three elements overlap due to the similar masses of the nuclei. Our results for SD capture in the Earth differ from the recent findings in Ref.~\cite{catena2016}, since those results are obtained using only the 11 most abundant elements on Earth as given in the {\tt DarkSUSY} package~\cite{gondolo2004} and Ref.~\cite{lundberg2004}, neglecting $^{25}$Mg, $^{29}$Si, and $^{55}$Mn. 

For the Sun, hydrogen dominates the SD capture rate~\cite{kappl2011}, however, $^{14}$N also contributes to SD capture, becoming important for $m_\chi \gtrsim 10\,$TeV.

\subsection{Are capture processes in equilibrium?} \label{Are the Sun or the Earth in equilibrium?}

The WIMP capture and annihilation processes reach equilibrium for times $t > \tann$, where the equilibrium time scale $\tann$ (cf. Eq.~\eqref{relaxation_time}) can be written in terms of the self-annihilation cross section $\sann$, the WIMP-proton cross section $\sigma_p^s$, and the local WIMP energy density $\rho^{\rm loc}_\chi$ as
\begin{equation} \label{tau_relaxation}
\tann = \left[\frac{K^{s}(m_\chi)}{\Ve}\,\sigma_p^{s}\,\rho^{\rm loc}_\chi\,\sann\right]^{-1/2}.
\end{equation}
If $\tann$ is greater than the age of the capturing body, for which we use the age of the solar system $t_\odot$ as a proxy, we refer to the processes as ``in equilibrium''. For $\tann > t_\odot$ we consider the processes being ``out of equilibrium''. As discussed below in Sec.~\ref{Subdominant WIMP DM model}, the self-annihilation cross section $\sann$ is fixed by requiring that the DM fraction in WIMPs is obtained through a thermal freeze-out mechanism.

The corresponding regions in the WIMP parameter space are shown by the solid black curves in Fig.~\ref{AnnRatePlotEarth} for the Earth and Fig.~\ref{AnnRatePlotSun} for the Sun for both SI and SD scattering. For scattering cross sections larger than those indicated by the solid curve, the processes are in equilibrium, while for smaller cross sections capture and annihilation have not yet reached equilibrium in the body. Current bounds from direct detection experiments rule out scattering cross sections large enough for the Earth to be in equilibrium for both SI and SD scattering. For the Sun, large enough scattering cross-sections for capture and annihilation to have reached equilibrium are not ruled out yet by direct detection. For capture via SD scattering, direct detection bounds are roughly four orders of magnitude weaker than the smallest cross sections required to be in equilibrium. For SI scattering, large enough scattering cross sections are marginally excluded for WIMP masses $m_\chi\approx 30\,$GeV, where the direct detection bounds from liquid Xe experiments are strongest, while for both smaller and larger WIMP masses sufficiently large cross sections are still allowed. The equilibrium time scales presented agree with the computation in Refs.~\cite{peter2009a, peter2009b}, which is performed for $m_\chi = 100\,$GeV.

\subsection{Sub-dominant WIMP DM model} \label{Subdominant WIMP DM model}

It is possible that WIMPs only make up a fraction $\fc$ of the total DM budget,
\begin{equation} \label{subdominant_energy_density}
\rho_\chi = \fc\,\rho_{\rm DM},
\end{equation}
where $\rho_\chi$ is the present cosmological abundance of WIMPs and $\rho_{\rm DM}$ is the present DM energy density. Current measurements of the cosmic microwave background constrain the DM budget $\Omega_{\rm DM}~=~\rho_{\rm DM}/\rho_{\rm crit}$ in terms of the critical energy density $\rho_{\rm crit} = 3H_0^2/8\pi\,G$ as~\cite{komatsu2009, ade2015}
\begin{equation} \label{DM_measured}
\Omega_{\rm DM}\,h^2 =  0.1199 \pm 0.0022,
\end{equation}
where $h = H_0/(100{\rm~km\,s^{-1}\,Mpc^{-1}})$ is the reduced Hubble constant and $H_0$ the present value of the Hubble rate. In the following, we consider the possibility that WIMPs make up only a fraction $\fc < 100\,\%$ of the DM, while the remaining DM, e.g. axions, may not get trapped in the Sun and Earth due to its light mass and/or small cross section~\cite{duda2002, duda2003}.

We assume the local WIMP energy density $\rho^{\rm loc}_\chi$ to scale with the global WIMP density
\begin{equation}
\rho^{\rm loc}_\chi = \fc\,\rho^{\rm loc}_{\rm DM},
\end{equation}
with the total local DM density given in Eq.~\eqref{DM_local} as $\rho^{\rm loc}_{\rm DM} = 0.4\,$GeV/cm$^3$. Bounds on the WIMP scattering cross section from direct detection are directly proportional to $\rho^{\rm loc}_\chi$ and hence are loosened as $\propto \fc^{-1}$.

\begin{figure}
\includegraphics[width=1\linewidth]{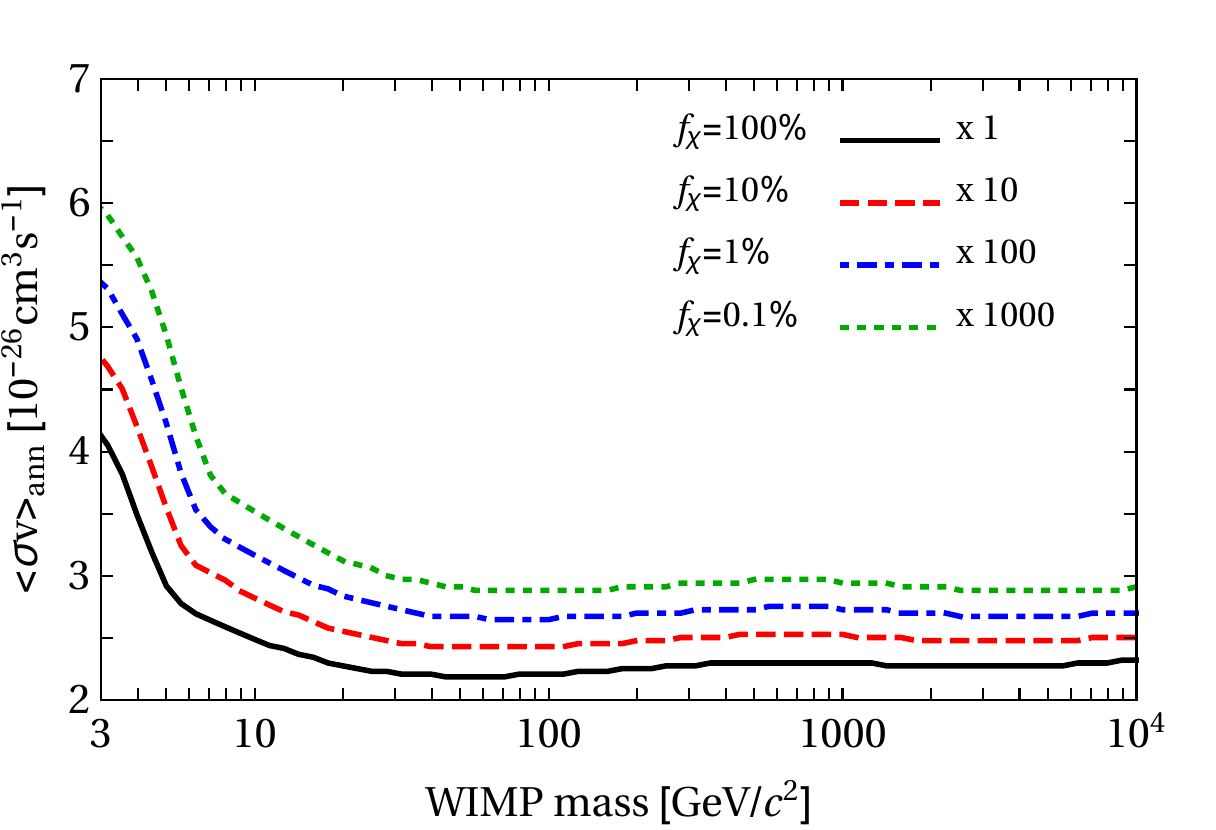}
\caption{The required thermally averaged annihilation cross section $\sann$ as a function of WIMP mass $m_\chi$ to get a WIMP abundance of $\fc = 100, 10, 1, 0.1\,\%$ assuming the standard freeze-out production mechanism, $s$-wave annihilation and no significant contribution from co-annihilation. Note that the lines are rescaled by the respective $\fc^{-1}$, such that they would end up exactly on top of each other if $\sann$ would scale as $\fc^{-1}$. The deviations from this scaling are caused by the change in the effective number of degrees of freedom at the time of decoupling, see Appendix \ref{appendix_relic} for a discussion. For example, when comparing the product $\fc \sann$ for $\fc = 1\,\%$ to the $\fc = 100\,\%$ case, the product increases by a factor of $\approx 1.2$.} 
\label{fig:relDens_1}
\end{figure}

Fixing the value of $\fc$ gives a precise relation between $\sann$ and the WIMP mass $m_\chi$ when assuming thermal freeze-out production, as we review in Appendix~\ref{appendix_relic}. In this work, we assume $s$-wave annihilation and no significant contribution from co-annihilation, thus, $\sann$ has the same numerical value in the early universe and today. Note that including $p$-wave and/or co-annihilation is straightforward in our framework. We show the required thermally averaged annihilation cross section as a function of WIMP mass $m_\chi$ to get a WIMP abundance of $\fc = 100,\ 10,\ 1,$ and $0.1\,\%$ in Fig.~\ref{fig:relDens_1}. The velocity-averaged annihilation cross section approximately scales as $1/\fc$, see the caption of Fig.~\ref{fig:relDens_1} and Appendix \ref{appendix_relic} for more details. Since $\tann \propto \left( \sigma_p^s \sann \rho_\chi^{\rm loc} \right)^{-1/2}$ is only mildly dependent on $f_\chi$, the region where the capture and the annihilation processes are in equilibrium is almost unaltered by a change in $\fc$ . For example, we find $\left[\sann \rho_\chi^{\rm loc}\right]_{\fc = 1\,\%}/\left[\sann \rho_\chi^{\rm loc}\right]_{\fc = 100\,\%} \approx 1.2$, so that the capture-annihilation equilibrium line moves down by a factor $\approx 1/1.2$ when comparing the two cases $\fc = 100\,\%$ and $\fc = 1\,\%$ in Figs.~\ref{muon_flux_wwearth}-\ref{muon_flux_bbsun}. See also Appendix~\ref{appendix_relic} and Refs.~\cite{duda2002, duda2003, gelmini2005, blum2014} for further discussion.

Previous work on capture for sub-dominant WIMP DM~\cite{duda2002} assumed a model-dependent relation between the scattering and the annihilation cross sections. Here, we treat the annihilation and scattering cross sections as independent quantities, since we do not consider a particular underlying model. Thus, the annihilation rate is a function of $\Gamma_A = \Gamma_A\left(m_\chi, \sigma^{s}_i, \sann\right)$, with $\sann$ determined by $\fc$. Given the age of the capturing body $t_{\odot}$, the value of the annihilation rate today is given by Eq.~\eqref{annihilation_rate_time},
\begin{equation}\label{regimes_annihilationrate} \begin{split}
\Gamma_A &= \frac{C}{2}\,\tanh^2\left(\frac{t_{\odot}}{\tann}\right) = \frac{K^{s}(m_\chi)}{2}\,\sigma_p^{s}\,\rho^{\rm loc}_\chi \times
\\ & \times \tanh^2\left\{\left[\frac{K^{s}(m_\chi)}{\Ve}\,\sigma_p^{s}\,\rho^{\rm loc}_\chi\,\sann\right]^{1/2}\,t_{\odot}\right\}. \end{split}
\end{equation}
This relation has different limiting behavior for the body being in and out of equilibrium
\begin{equation} \label{regimes_annihilationrate_limits}
\Gamma_A \approx
\begin{cases}
\frac{\sann\,t_{\odot}^2}{2\,\Ve}\,\left[K^{s}(m_\chi)\,\sigma_p^{s}\,\rho^{\rm loc}_\chi\right]^2 & \hbox{for $t_{\odot} \lesssim \tann$},\\
\frac{K^{s}(m_\chi)}{2}\,\sigma_p^{s}\,\rho^{\rm loc}_\chi  & \hbox{for $t_{\odot} \gtrsim \tann$}.
\end{cases}
\end{equation}
For given values of $\sigma_p^{s}$ and $m_\chi$, $\Gamma_A$ scales nearly linearly with $\fc$ in both regimes, since the product $\sann\,\rho^{\rm loc}_\chi$ is approximately constant in $\fc$. The dependence of $\Gamma_A$ on $\sigma^{s}_p$ is quadratic when the capture process is out of equilibrium and linear when in equilibrium. 

\begin{figure*}
\includegraphics[width=\linewidth]{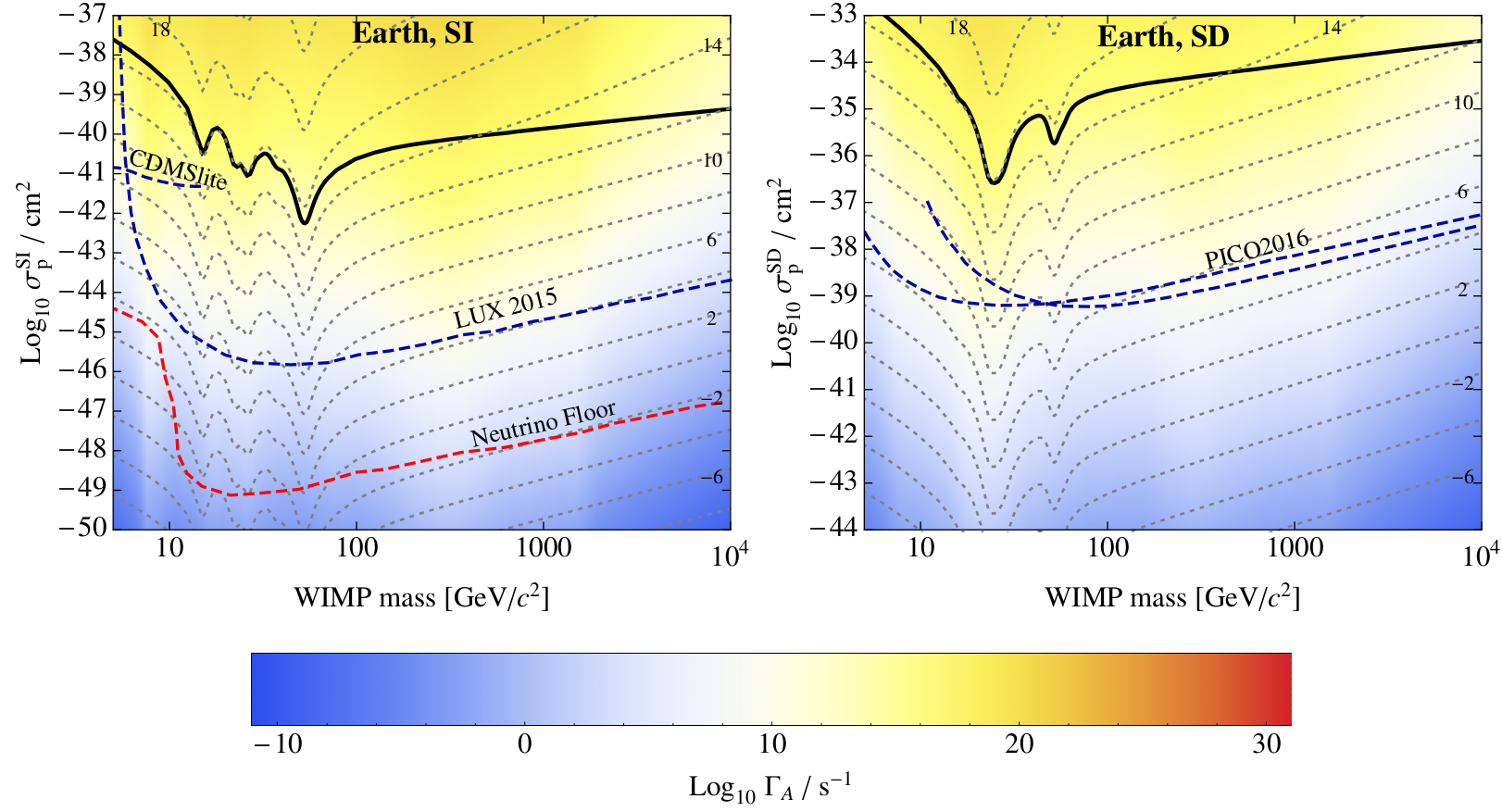}
\caption{The color scale gives the value of $\log_{10}(\Gamma_A/{\rm s}^{-1})$ as a function of the WIMP mass $m_\chi$ (on the X-axis) and of the WIMP-proton scattering cross section (on the Y-axis), assuming that WIMPs make up $\fc=100\,\%$ of DM. Also shown are contour lines (dotted gray) for specific values of $\log_{10}(\Gamma_A/{\rm s}^{-1})$ equal to the number labelling the contour. The solid black line represents the boundary of the region where the Earth has reached equilibrium between capture and annihilation. Above this line, the Earth is in equilibrium while below it has not reached equilibrium yet. We have included current bounds (dashed blue) from CDMS~\cite{agnese2014} and LUX~\cite{akerib2014} (SI) and PICO~\cite{amole2016} (SD), plus the expected neutrino floor to be detected in future direct detection experiments (dashed red). For sub-dominant WIMPs, the annihilation rate scales approximately as $\fc$ while the boundary of the region where capture and annihilation rates are in equilibrium remains approximately unchanged since $\tann \propto \left(\sigma^s_p \sann \fc \right)^{-1/2}$. Deviations from this approximate behaviour are induced by the deviations of the scaling of $\sann$ from $\sann \propto \fc^{-1}$, as discussed in the text and in Appendix \ref{appendix_relic}, and would not be visible by eye on the scales shown in this Figure, cf. Fig \ref{muon_flux_wwearth}.}
\label{AnnRatePlotEarth}
\end{figure*}

We show the value of $\Gamma_A$ (color scale) as a function of $\sigma_p^{s}$ and $m_\chi$, for WIMPs constituting $\fc = 100\,\%$ of the DM. We show panels for SI (left) and SD (right) scattering for the Earth in Fig.~\ref{AnnRatePlotEarth} and the Sun in Fig.~\ref{AnnRatePlotSun} . The gray dashed lines represent curves where the annihilation rate is constant. We label these lines by the exponent of the annihilation rate $\zeta = \log_{10}\left(\Gamma_A/{\rm s}^{-1}\right)$. The black line is the boundary between capture and annihilation being in or out of equilibrium, as discussed in the previous subsection. For each figure, the different spacing between the gray curves above and below the black line reflects the change in the scaling of the annihilation rate on $\sigma_p^{s}$ in the two different regimes given in Eq.~\eqref{regimes_annihilationrate_limits}. We remark that fixing $\fc$ gives a unique choice of $\sann$ as a function of $m_\chi$, such that the annihilation cross section is not a free parameter of the plot.

When the contribution of WIMPs to the total DM energy density is smaller, we expect the annihilation rate to scale approximately as $\fc$ in the whole parameter space, as seen in Eq.~\eqref{regimes_annihilationrate_limits}. For example, a subdominant WIMP model with $\fc =1\%$ will have a $\Gamma_A$ approximately $100$ times smaller than a model with $\fc =100\%$.

\section{Muon flux at Super-K and IceCube} \label{Muon flux at Super-K and IceCube}

\begin{figure*}
\includegraphics[width=\linewidth]{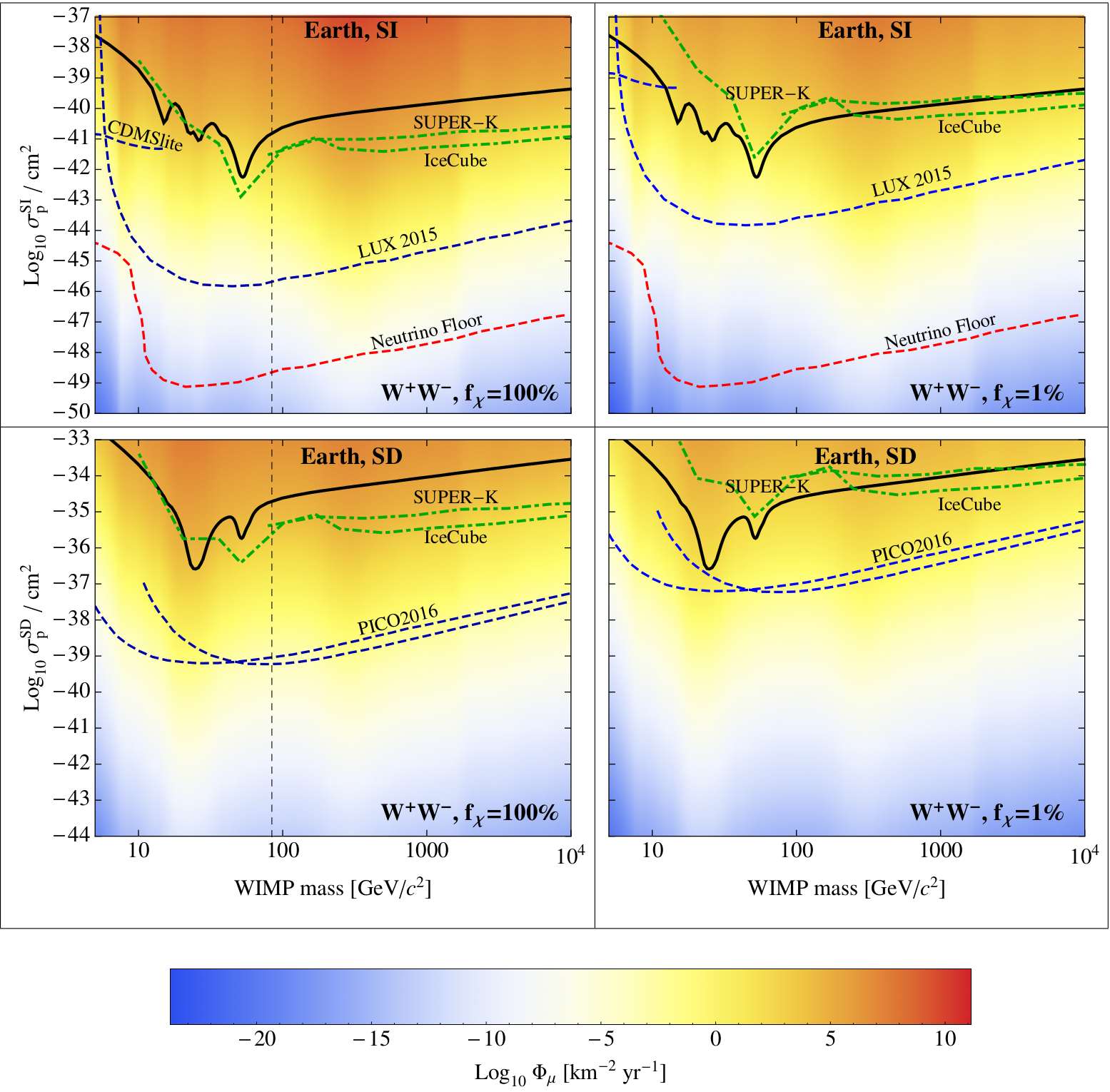}
\caption{Muon flux at the detector (in km$^{-2}$\,yr$^{-1}$) for capture in the Earth via SI (upper panels) and SD scattering (lower panels) and WIMP annihilation into ${\rm W^+W^-}$. We assume a WIMP fraction $\fc = 100\%$ (left) or $\fc = 1\%$ (right) of the DM. Current upper bounds from Super-K and IceCube are shown in dashed-dotted green lines. We also show the region where the capture rate is out of equilibrium (black lines), current bounds from CDMS, LUX, and PICO (dashed blue lines), and the neutrino floor (dashed red lines), as shown in Fig.~\ref{AnnRatePlotEarth}.  For the case
of  $\fc=100\%$, the region to the left of the dashed vertical line is ruled out by MAGIC and Fermi-LAT measurements. For $\fc = 1\,\%$ the MAGIC/Fermi-LAT bound rules out WIMP masses smaller than shown in these Figures.}
\label{muon_flux_wwearth}
\end{figure*}

\begin{figure*}
\includegraphics[width=\linewidth]{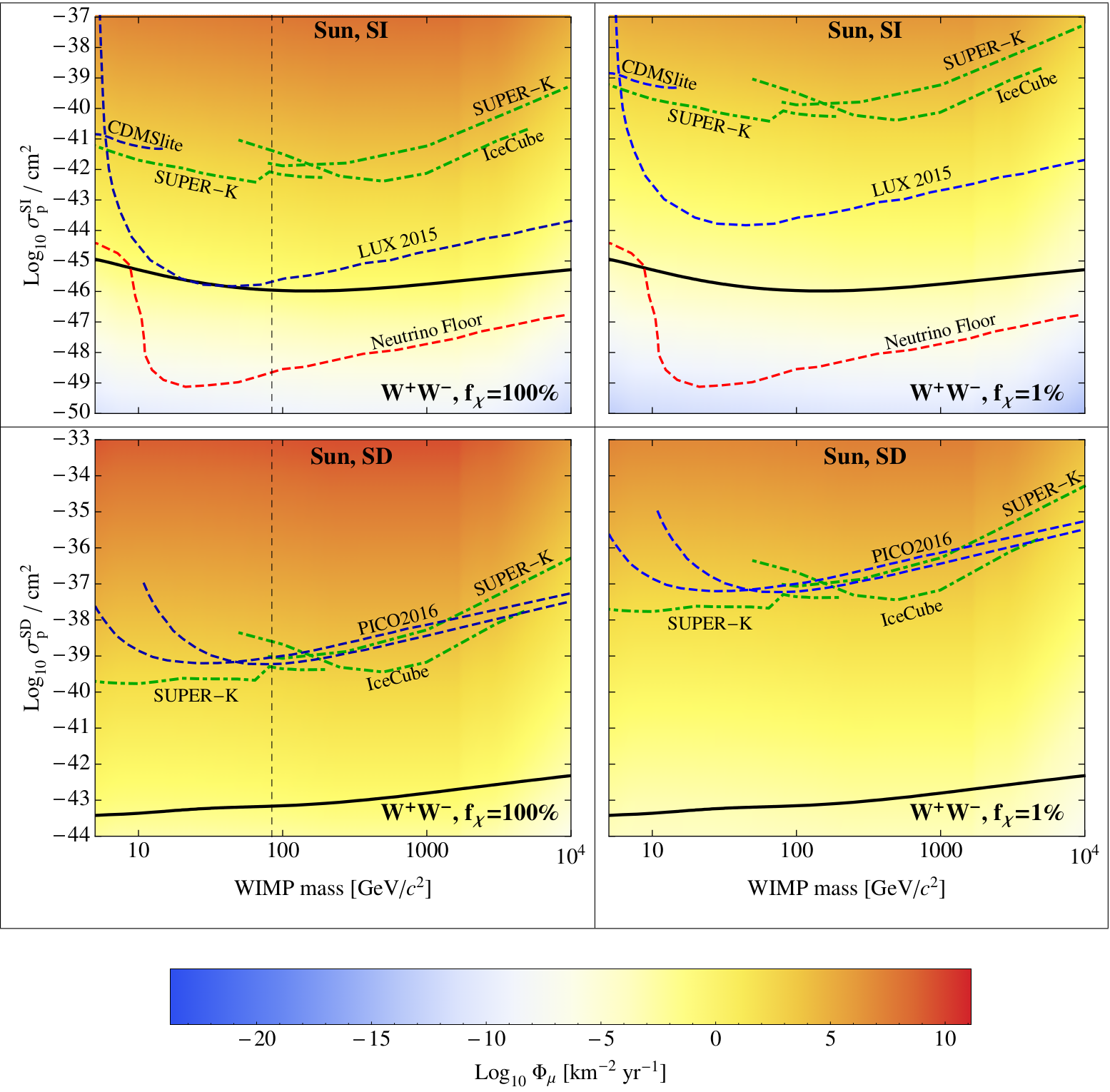}
\caption{Same as Fig.~\ref{muon_flux_wwearth}, but for solar capture, with the direct detection bounds and the neutrino floor as shown in Fig.~\ref{AnnRatePlotSun}.}
\label{muon_flux_wwsun}
\end{figure*}

\begin{figure*}
\includegraphics[width=\linewidth]{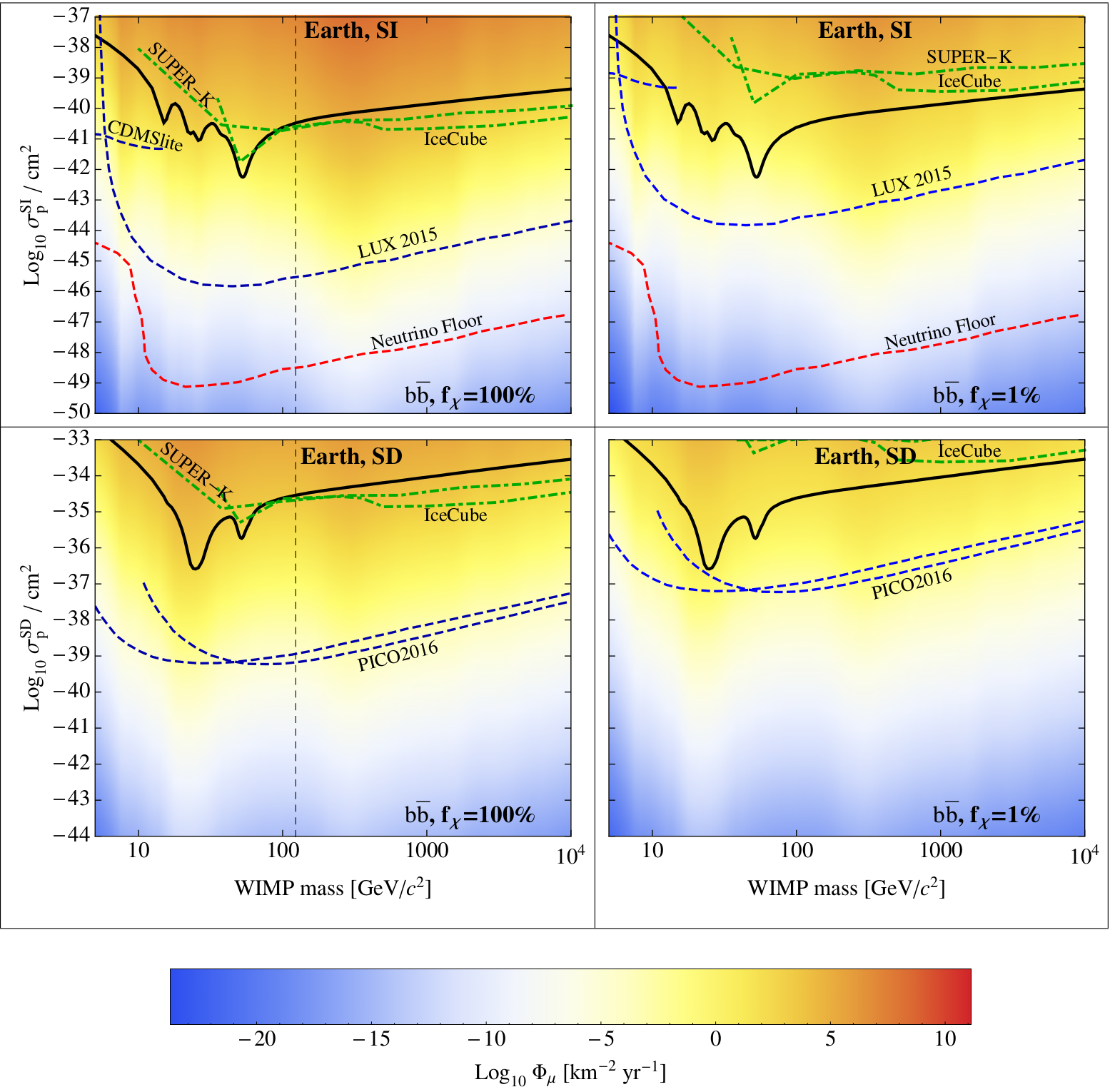}
\caption{Same as Fig.~\ref{muon_flux_wwearth}, but for Earth capture and WIMP annihilation into ${\rm b\bar{b}}$.}
\label{muon_flux_bbearth}
\end{figure*}

\begin{figure*}
\includegraphics[width=\linewidth]{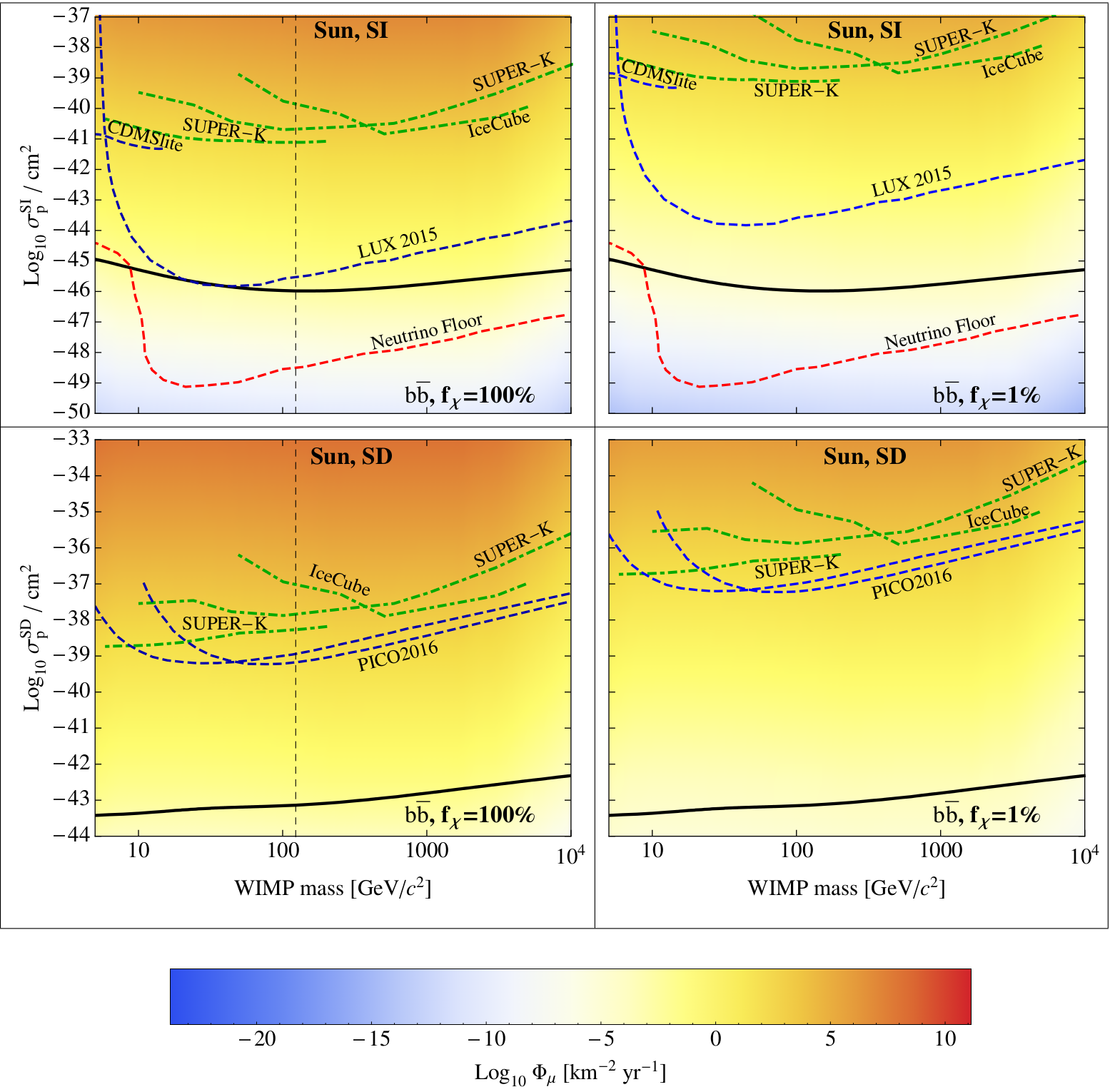}
\caption{Same as Fig.~\ref{muon_flux_wwsun}, but for solar capture and WIMP annihilation into ${\rm b\bar{b}}$.}
\label{muon_flux_bbsun}
\end{figure*}

WIMPs captured in Earth or the Sun annihilate into SM particles $X$ with WIMP-model dependent branching ratios $\mB_\chi^X = \mB_\chi(\chi \bar{\chi} \to X\bar{X})$. Unless stable, the primary decay product $X$ will then decay to lighter particles, eventually yielding photons, electrons, neutrinos, and the lightest hadrons. Of these particles, only neutrinos can travel freely through the capturing body and are thus the only product of the WIMP annihilation that will reach the surface of the Earth. There, they can be detected by neutrino observatories such as IceCube, Super-K, AMANDA, or ANTARES. However, such neutrino observatories do not detect neutrinos directly, but the Cherenkov light produced in the detector by muons from charged-current interactions of neutrinos inside or close to the detector. Hence, for our case of WIMP annihilation in the Earth or Sun, the quantity constrained by neutrino detectors is the muon-flux through the detector induced by the muon-neutrinos from the WIMP annihilations in the capturing body. The integrated muon-flux from WIMP annihilation at the detector is
\begin{equation} \label{muon_flux}
\Phi_\mu^{DM} = \Gamma_A \times  Y(m_\chi, \mB_\chi^X).
\end{equation}
The muon yield $Y$ per area and WIMP-annihilation is given by (cf.~\cite{abbasi2011, catena2016})
\begin{equation} \label{yield_eq}
\begin{split} Y(m_\chi, \mB_\chi^X) &= n_T \int dE_\mu \, \int d\lambda \, \frac{d\mP(E_\mu, E_\mu'; \lambda)}{dE_\mu\,d\lambda} \times
\\ & \times \int dE_\mu' \int \frac{dE_\nu}{4\pi D^2} \, \frac{d\sigma_T(E_\mu', E_\nu)}{dE_\mu'} \times
\\ & \times \sum_{\ell'} \mP_{\nu_{\ell'}\to\nu_\mu}(E_\nu, D)\,\sum_{X} \mB_\chi^X \frac{d\Nni}{dE_\nu}, \end{split}
\end{equation}
where $ d\sigma_T(E_\mu', E_\nu)/dE_\mu'$ is the differential charged-current cross section for production of a muon with energy $E_\mu'$ by a neutrino scattering of target nuclei with a number density $n_t$, and $d\mP(E_\mu, E_\mu'; \lambda)/dE_\mu d\lambda$ is the probability per energy and length to find a muon with energy $E_\mu$ in the detector after it travelled a distance $\lambda$.

In the previous sections we computed $\Gamma_A$. The muon yield $Y$ is usually obtained by performing a Monte Carlo simulation over the decay chains of the primary WIMP annihilation products $X$, the propagation of the resulting neutrinos from the production site to the detector including oscillations, and finally the interactions of the neutrinos at the detector site producing muons and their propagation into the detector. We use the results for $Y$ from {\tt WimpSim}~\cite{Blennow2007}, which performs such a Monte Carlo simulation for both the Earth and the Sun including matter effects, tabulated for a range of WIMP masses and primary decay channels.

The muon flux from DM annihilation has been constrained by Super-K for capture by the Earth~\cite{tanaka2011b} and in the Sun~\cite{tanaka2011a, tanaka2011b, choi2015a, choi2015b}\footnote{The strongest bounds for low-mass WIMPs in the Sun from Super-K~\cite{choi2015a,choi2015b} have been presented as upper limits on the muon-neutrino flux in Refs.~\cite{choi2015a,choi2015b}. We convert this to an upper limit on the muon flux by rescaling these limits with the ratio of the muon yield $Y$ and the corresponding quantity for the muon-neutrino flux at the detector from WIMP annihilations in the Sun.}. IceCube is sensitive to neutrinos with higher energy and thus constrains the flux for larger WIMP masses for the Earth~\cite{aartsen2016b} and  the Sun~\cite{abbasi2011, aartsen2016a}. See Ref.~\cite{zornoza2016} for recent results at the AMANDA telescope. Neutrino observatories usually present their bounds assuming annihilation into one channel $X$ at a time, $\mB_\chi^X = 1$. To compare our results with the bounds from IceCube and Super-K, we present them assuming annihilation to ${\rm b\bar{b}}$ only, which gives particularly soft neutrino spectra, and to ${\rm W^+W^-}$ only, which yields harder spectra. For WIMPs lighter than $W$-bosons, we assume annihilation to $\tau^+\tau^-$ for the hard channel. WIMPs captured in the Earth or  the Sun are non-relativistic and hence for $m_\chi < m_W$ the $\chi \bar{\chi} \to {\rm W^+W^-}$ decay is kinematically suppressed. $\tau$ decay chains yield hard neutrino spectra and thus take the role of the $\chi \bar{\chi} \to {\rm W^+W^-}$ channel for light WIMPs.

We show results in Fig.~\ref{muon_flux_wwearth} (\ref{muon_flux_bbearth}) for muons originating from Earth capture and Fig.~\ref{muon_flux_wwsun} (\ref{muon_flux_bbsun}) for muons originating from solar capture together with the relevant bounds, assuming WIMPs to annihilate to ${\rm W^+W^-}$ (${\rm \bar{b}b}$). We have converted the bounds on the muon flux to bounds over the SI and SD WIMP-proton cross sections using Eq.~\eqref{muon_flux}, with the annihilation rate in Eq.~\eqref{regimes_annihilationrate} and the yield in Eq.~\eqref{yield_eq}. Neutrino observatories rule out regions of the parameter space with scattering cross sections smaller than those required for the Earth to have reached equilibrium yet, but are less constraining than direct detection bounds on SI and SD scattering.

\subsection{Scaling relationships}\label{sec:scaling}

We discuss in depth the scaling relationship of different quantities with $\fc$ in different regions of the plots in Figs.~\ref{muon_flux_wwearth}-\ref{muon_flux_bbsun} Although such results can easily be derived, they have not been discussed in previous literature, except to some extent in Refs.~\cite{duda2002, duda2003} for supersymmetric WIMP models. According to Eq.~\eqref{muon_flux}, the muon flux from annihilation in the Earth or Sun
follows the same scaling relation as that for $\Gamma_A$ in Eq.~\eqref{regimes_annihilationrate_limits} which, when re-written as a function of $\fc$, reads
\begin{equation} \label{regimes_flux_limits}
\Phi_\mu \propto
\begin{cases}
\sann \left(\sigma^s_p\right)^2 \fc^2 \propto \left(\sigma^s_p\right)^2 \fc & \hbox{for $t_{\odot} \lesssim \tann$},\\
\sigma_p^{s}\,\fc  & \hbox{for $t_{\odot} \gtrsim \tann$}.
\end{cases}
\end{equation}
In the first line, we have used the fact that appoximately $\sann \propto \fc^{-1}$ (see Fig.~\ref{fig:relDens_1}). The bounds on the WIMP-proton scattering cross section due to indirect searches from the Earth and Sun at a given muon flux thus scale as
\begin{equation} \label{regimes_sigma_limits}
\sigma^s_p \propto
\begin{cases}
\fc^{-1/2} & \hbox{for $t_{\odot} \lesssim \tann$},\\
\fc^{-1}  & \hbox{for $t_{\odot} \gtrsim \tann$},
\end{cases}
\end{equation}
while the bounds from direct detection scale as $\sigma^s_p \propto \fc^{-1}$ in the whole region of the parameter space. As an example, consider the bound from IceCube for the Earth in Fig.~\ref{muon_flux_wwearth}. Since this bound is placed in the region $t_{\odot} \lesssim \tann$, outside the equilibrium region of the Earth, the green dashed line moves up by one order of magnitude when $\fc$ changes from $100\%$ to $1\%$. On the contrary, the IceCube bound for the Sun in Fig.~\ref{muon_flux_wwsun} moves up by two orders of magnitude when $\fc$ changes from $100\%$ to $1\%$, since the bound is placed in the region $t_{\odot} \gtrsim \tann$ where capture and annihilation are in equilibrium. Regardless of the capture and annihilation processes, the direct detections bounds from LUX and PICO become weaker by two orders of magnitude when $\fc$ changes from $100\%$ to $1\%$, for both the Earth and the Sun.

Neutrino observatories place bounds that might be competitive with the direct detection measurements. Considering $\chi\bar{\chi} \to {\rm W^+W^-}$, Fig.~\ref{muon_flux_wwsun}, for SD capture in the Sun it is current bounds from neutrino observatories that put stronger limits on the WIMP-proton cross section than direct detection experiments, while for SI capture it is direct detection which provides the most stringent constraints. This holds both for WIMPs composing all of DM $\fc = 100\,\%$ and for sub-dominant WIMP DM, e.g. for the $\fc = 1\,\%$ case we show in our plots.

\enlargethispage{\baselineskip}
For the case of $\chi\bar{\chi} \to {\rm b\bar{b}}$, the same discussion of the scaling of the bounds on the scattering cross section with $\fc$ also applies, see Figs.~\ref{muon_flux_bbearth}-\ref{muon_flux_bbsun}. However, since the neutrino spectra are softer, the bounds are somewhat weaker than the respective bounds assuming $\chi \bar{\chi} \to {\rm W^+W^-}/\tau^+\tau^-$ annihilation. Also note that the vertical dashed line in the plots with $\fc = 100\,\%$ shows the most recent MAGIC/Fermi-LAT constraint from WIMP annihilation yielding photons in dwarf satellite galaxies~\cite{Ahnen:2016qkx}. The region to the left of the dotted line is ruled out. In general, for both the hard and soft annihilation channels we consider, the thermal relic annihilation cross section is ruled out for $m_\chi \lesssim 100\,$GeV, assuming $\fc = 100\,\%$. These bounds are subject to large astrophysical uncertainties~\cite{geringer2011, geringer2012, ullio2016}, which could considerably weaken or strengthen such constraints. For models with sub-dominant WIMP densities, the MAGIC/Fermi-LAT constraints are weakened since indirect detection bounds on the WIMP annihilation cross section scales $\propto \fc^{-2}$ and the relevant thermal relic cross section only scales $\propto \fc^{-1}$. Current bounds from MAGIC/Fermi-LAT rule out WIMPs of mass $m_\chi \lesssim 1\,$GeV for $\fc = 1\,\%$, below the mass range we consider. Neutrinos in IceCube/DeepCore coming from the direction of the Galactic Center or dwarf spheroidal galaxies  can also be used to set limits on $\sann$~\cite{leptophilic,sandick2010} for the case of leptophilic DM. However, such results are currently too weak to set bounds in the parameter space we show.

\subsection{Effect of updated composition of the Earth} \label{Effect of updated composition of the Earth}

\begin{figure}
\includegraphics[width=1\linewidth]{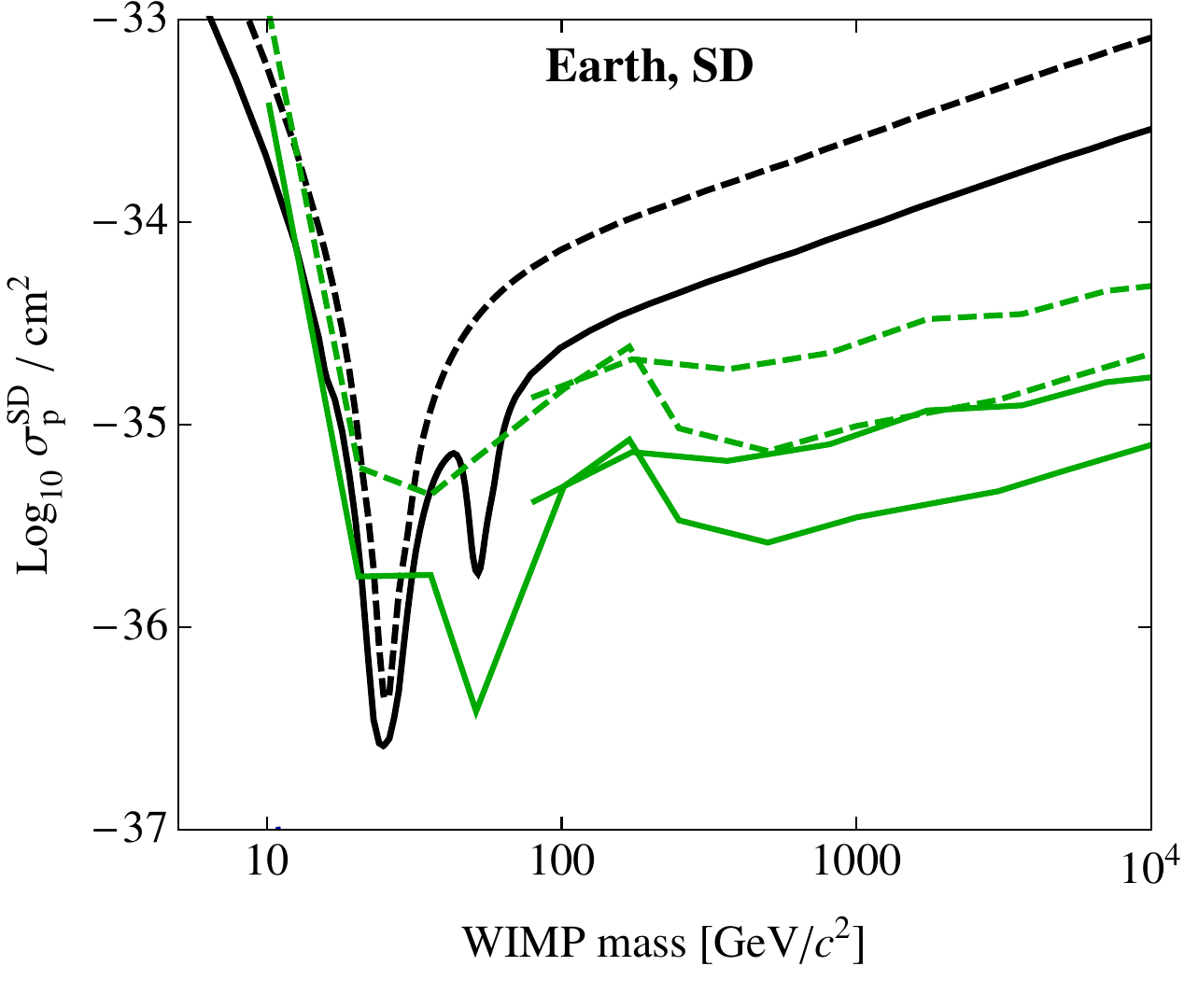}
\caption{The curve $\tann = t_{\odot}$ for SD capture in the Earth, considering the elements in {\tt DarkSUSY} (dashed black line) and the set of nuclei used in this paper (solid black line). Also shown are the corresponding bounds from the muon flux. The peak at $m_\chi\sim 25\,$GeV is due to $^{25}$Mg, $^{27}$Al, and $^{29}$Si, while the less pronounced peak at $m_\chi\sim 52\,$GeV is due to $^{55}$Mn.}
\label{ComparisonPlot_label}
\end{figure}

As discussed in Sec.~\ref{Dark matter capture rate by a massive body}, we have updated the chemical composition of the Earth used to compute the WIMP capture process in the Earth with respect to the composition tabulated in {\tt DarkSUSY} and used in the recent literature on DM capture~\cite{catena2016}. While the impact of the updated composition is negligible for capture via SI scattering, for SD capture we find an increase of the capture rate and thus also the muon flux by more than a factor three with respect to using the composition of the Earth as tabulated in \texttt{DarkSUSY}. We compare results in Fig.~\ref{ComparisonPlot_label}. The solid black line shows the $\tann = t_{\odot}$ curve for our updated Earth composition, while the same curve obtained with the elements in {\tt DarkSUSY} is shown in the dashed black line. Also shown are the bounds from the muon flux discussed in Sec.~\ref{Muon flux at Super-K and IceCube} for updated (solid green) and {\tt DarkSUSY} (dashed green) chemical compositions of the Earth. Updating the Earth abundances improves bounds on $\sigma_p^{\rm SD}$ by approximately a factor three. The new peak at $m_\chi \sim 52\,$GeV, appearing in the solid black line, is due to our inclusion of $^{55}$Mn in the computation of the capture rate, while the difference between the height of the two peaks at $m_\chi\sim 30$GeV is due to the inclusion of $^{25}$Mg and $^{29}$Si. Although bounds on $\sigma_p^{\rm SD}$ improve by a factor three, the capture rate for the Earth for both SI and SD scattering is too low to provide bounds competitive with current direct-detection limits, as shown in Figs.~\ref{muon_flux_wwearth} and~\ref{muon_flux_bbearth}.

\section{Conclusion}

Neutrinos from annihilation of WIMP DM captured in massive bodies such as the Sun and the Earth provide a complementary test of WIMP models to direct detection, as well as to other indirect detection methods. The signal can be sensitive to both the WIMP-nuclei scattering cross section through which the capture itself proceeds and the WIMP annihilation cross section giving rise to the neutrino flux. Direct detection experiments on the other hand are sensitive to the WIMP-nuclei scattering cross section only, while other indirect detection searches depend only on the annihilation cross section.

In this work, we have considered the standard cases of SI and SD scattering only, see Refs.~\cite{catena2015, catena2016} for capture in the non-relativistic EFT DM framework. Comparing current bounds from the Super-K and IceCube neutrino observatories with direct detection bounds from LUX, CDMSLite and PICO we find that direct detection places stronger bounds on the SI scattering cross-section, but that neutrinos from capture in the Sun give the strongest bound on the SD scattering cross section excluding $\SDp \gtrsim 10^{-40}\,$cm$^2$ for $m_\chi \lesssim 1\,$TeV. However, even where direct detection bounds are stronger, bounds from WIMP capture and annihilation provide an important check due to different systematic uncertainties. For example, direct detection experiments often rely on one target element only whereas capture in the Sun or the Earth proceeds via scattering off a number of different elements.

We used a refined model for the composition of the Earth, finding that the bounds on the SD cross section from the measured muon flux at Super-K and IceCube are strengthened by approximately a factor three compared to previous results, see Sec.~\ref{Effect of updated composition of the Earth}. This is due to our inclusion of additional elements responsible for SD capture, mainly $^{25}$Mg, $^{29}$Si, and $^{55}$Mn.

We have considered two scenarios: i) the case where WIMPs comprise the totality of the DM and ii) the case of sub-dominant WIMPs, in which they comprise a smaller fraction $\fc < 100\,\%$ of the total DM. Assuming thermal production, the annihilation cross section scales approximately as $\sann \propto \fc^{-1}$. Thus, bounds on the WIMP cross sections from direct detection scale as $\fc^{-1}$ and bounds from indirect detection as $\sann \fc^{-2} \propto \fc^{-1}$. The scaling of bounds from WIMPs captured in the Sun or the Earth depends on the equilibrium time scales as discussed in Sec. \ref{sec:scaling}. For the Sun, cross sections that can be ruled out by neutrino observatories firmly sit in the region where equilibrium is reached and bounds on the scattering cross section scale like those from direct and other indirect detection $\propto \fc^{-1}$. For the Earth on the other hand, neutrino observatories rule out scattering cross sections for which capture and annihilation have yet to reach equilibrium and bounds thus scale as $\propto\fc^{-1/2}$. Since direct detection bounds on the scattering cross section scale as $\fc^{-1}$, bounds from capture and annihilation in the Earth become more competitive with direct detection bounds for sub-dominant WIMP DM models.

\begin{acknowledgments}
We would like to thank Riccardo Catena, Joakim Edsj{\"o}, Paolo Gondolo, William F. McDonough, Sofia Sivertsson, and Axel Widmark for the useful discussions and comments that led to the present work. SB, KF, and LV acknowledge support by Katherine Freese through a grant from the Swedish Research Council (Contract No. 638-2013-8993). KF and PS acknowledge support from DoE grant DE-SC007859 at the University of Michigan.
\end{acknowledgments}

\newpage
\appendix

\onecolumngrid

\section{Table and Figures for Solar Capture} \label{Solar Capture}

\renewcommand{\thefigure}{A\arabic{figure}}

\setcounter{figure}{0}

\begin{table}[h!]
\begin{center}
\begin{tabular}{|c|c|c||c|c|c|}
\hline
Isotope & Mass fraction & Potential & Isotope & Mass fraction & Potential\\
$i$ & $x_i$ & $\phi_i$ & $i$ & $x_i$ & $\phi_i$\\
\hline
$^1$H & 0.684 & 3.15 & $^{24}$Mg & $7.30 \times 10^{-4}$ & 3.22\\
$^4$He & 0.298 & 3.40 & $^{27}$Al & $6.38 \times 10^{-5}$ & 3.22\\
$^3$He & $3.75 \times 10^{-4}$ & 3.40 & $^{28}$Si & $7.95 \times 10^{-4}$ & 3.22\\
$^{12}$C & $2.53 \times 10^{-3}$ & 2.85 & $^{32}$S & $5.48 \times 10^{-4}$ & 3.22\\
$^{14}$N & $1.56 \times 10^{-3}$ & 3.83 & $^{40}$Ar & $8.04 \times 10^{-5}$ & 3.22\\
$^{16}$O & $8.50 \times 10^{-3}$ & 3.25 & $^{40}$Ca & $7.33 \times 10^{-5}$ & 3.22\\
$^{20}$Ne & $1.92 \times 10^{-3}$ & 3.22 & $^{56}$Fe & $1.42 \times 10^{-3}$ & 3.22\\
$^{23}$Na & $3.94 \times 10^{-5}$ & 3.22 & $^{58}$Ni & $8.40 \times 10^{-5}$ & 3.22\\
\hline
\end{tabular}
\caption{The 16 most abundant isotopes of the Sun, their total mass fractions, and their effective gravitational potential $\phi_i$, as given in Ref.~\cite{gondolo2004}.}
\label{table_fraction_sun}
\end{center}
\end{table}

\begin{figure}[h!]
\includegraphics[width=\linewidth]{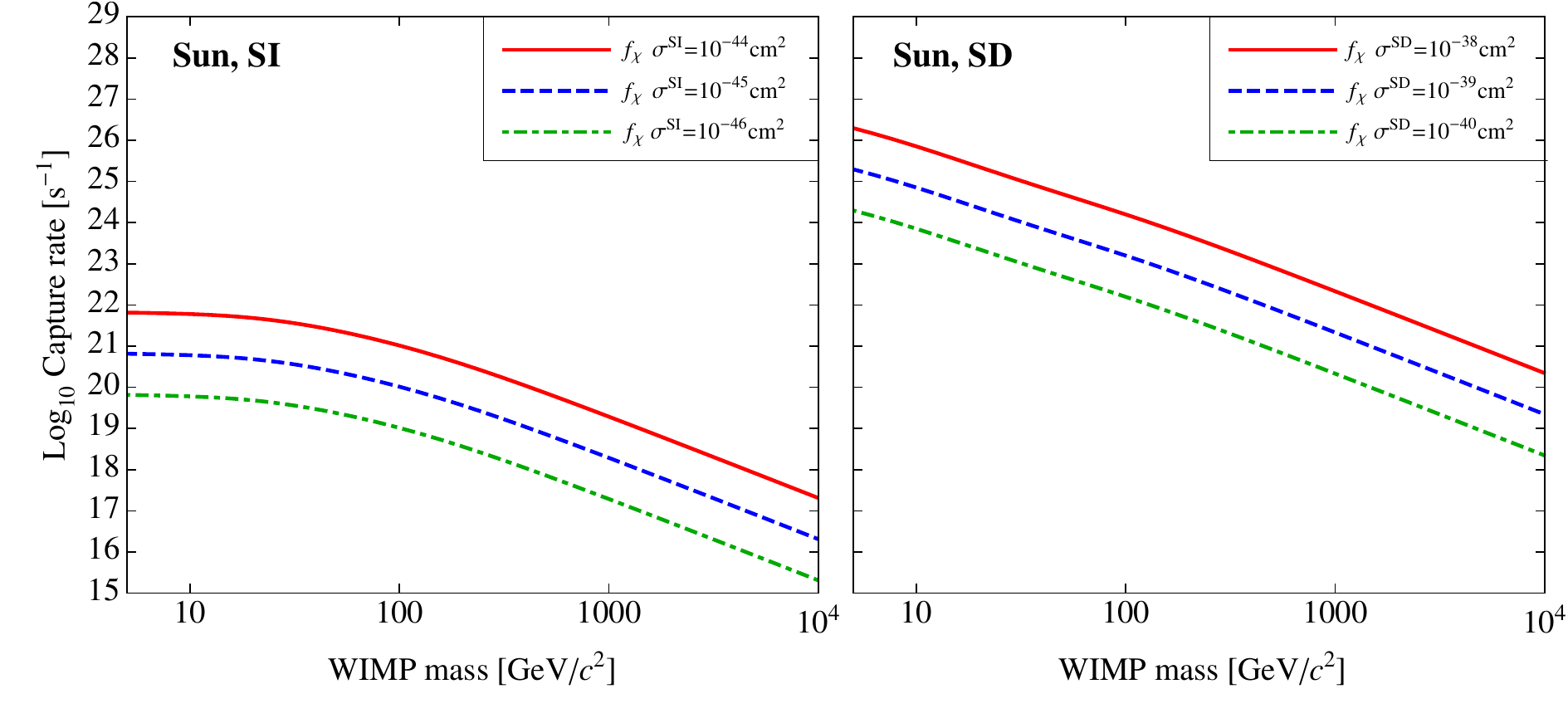}
\caption{The total capture rate (in s$^{-1}$) in the Sun, as a function of the DM mass~$m_\chi$, for different values of the WIMP-nucleon cross section. red: $\SI = 10^{-44}\,$cm$^2$ or $\SDp = 10^{-38}\,$cm$^2$; blue: $\SI = 10^{-45}\,$cm$^2$ or $\SDp = 10^{-39}\,$cm$^2$; yellow: $\SI = 10^{-46}\,$cm$^2$ or $\SDp = 10^{-40}\,$cm$^2$. The parameter $\fc$ is introduced later in the text and gives the WIMP fraction of the total DM budget.}
\label{fig_rate_sun}
\end{figure}

\begin{figure}[h!]
\includegraphics[width=\linewidth]{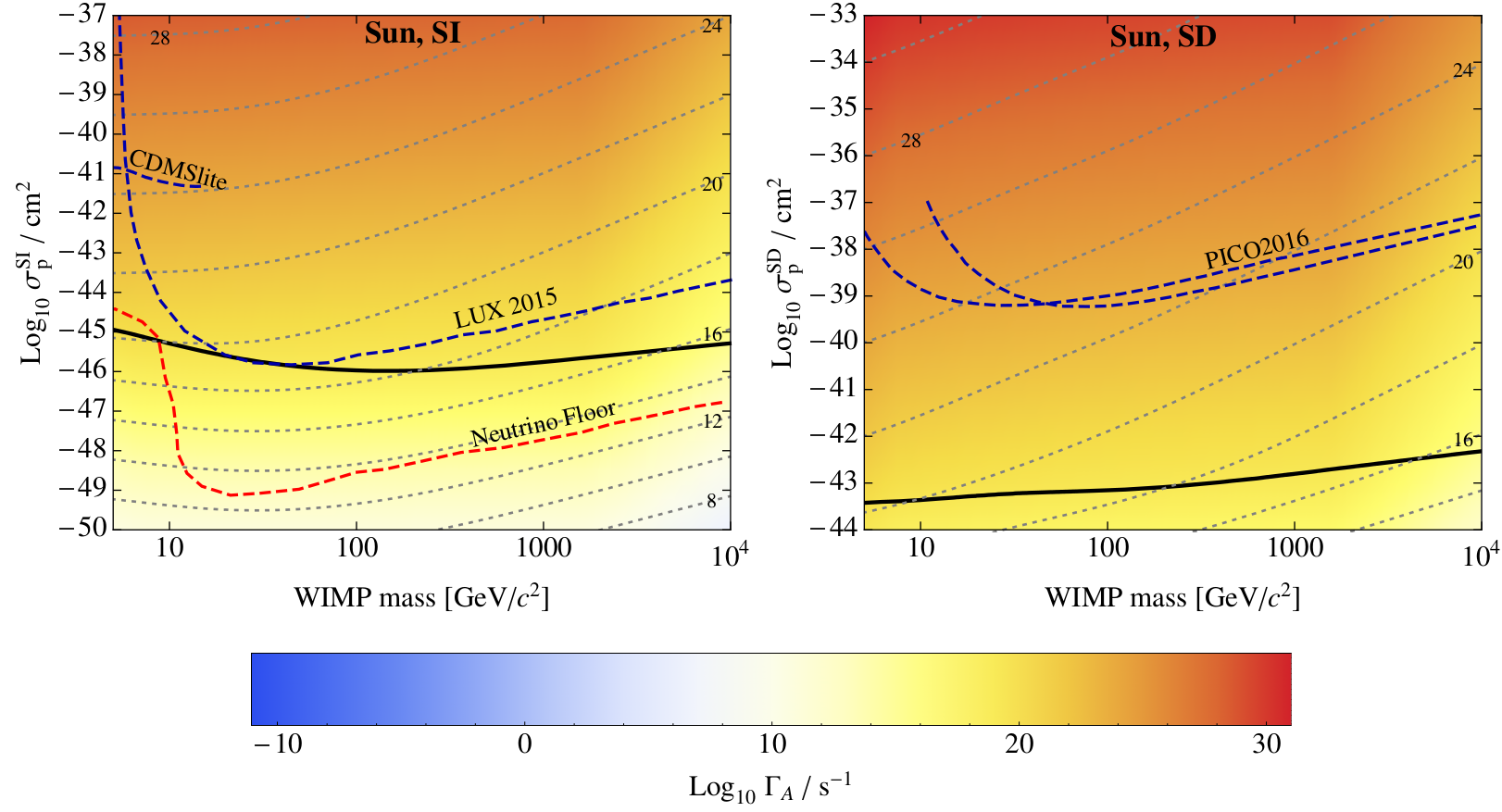}
\caption{The color scale gives the value of $\log_{10}(\Gamma_A/{\rm s}^{-1})$ as a function of the WIMP mass $m_\chi$ (on the X-axis) and of the WIMP-proton scattering cross section (on the Y-axis), assuming that WIMPs make up $\fc=100\,\%$ of DM. Also shown are contour lines (dotted gray) for specific values of $\log_{10}(\Gamma_A/{\rm s}^{-1})$ equal to the number labelling the contour. The solid black line represents the boundary of the region where the Sun has reached equilibrium between capture and annihilation. Above this line, the Sun is in equilibrium while below it has not reached equilibrium yet. We have included current bounds (dashed blue) from CDMS~\cite{agnese2014} and LUX~\cite{akerib2014} (SI) and PICO~\cite{amole2016} (SD), plus the expected neutrino floor to be detected in future direct detection experiments (dashed red).}
\label{AnnRatePlotSun}
\end{figure}

\twocolumngrid
\newpage
\section{Review of the annihilation rate} \label{annihilation_rate_rev}

The constant $C_A$ is obtained from Eqs.~\eqref{annihilation_rate0},~\eqref{number_density}, and~\eqref{annihilation_rate},
\begin{equation} \label{def_CA} \begin{split}
C_A &= 2\sann\,\int \,\tilde{n}^2(\br)\, d^3\br
\\ &= 2\sann\,\frac{\int d^3\br \,e^{-2m_\chi\,\Phi(r)/T}}{\left[\int d^3\br \, e^{-m_\chi\,\Phi(r)/T}\right]^2}. \end{split}
\end{equation}
In the instantaneous thermalization approximation, $C_A$ does not depend on time. For a constant density, we write the gravitational potential inside the body as
\begin{equation} \label{def_grav_potential}
\Phi(r) = \frac{2\pi}{3}\,G\,\bar{\rho}\, r^2 = \frac{T}{m_\chi}\,\frac{r^2}{r_\chi^2},
\end{equation}
where the thermal radius $r_\chi$, which describes the radius in which most of WIMPs are concentrated in, is given by
\begin{equation} \label{cutoff_radius}
r_\chi = \sqrt{\frac{3\,T}{2\pi\,G\,\bar{\rho}\,m_\chi}}.
\end{equation}
In principle, the WIMP temperature and the density profile depend on $r$, although for the Sun $T$ is well approximated by the core temperature for $m_\chi \gtrsim10\,$GeV~\cite{liang2016}. Using the temperature $T_{\odot} = 1.57\times 10^{7}\,$K and density $\bar{\rho}_{\odot} = 1.5\times 10^5\,$kg/m$^3$ of the Sun's core gives
\begin{equation} \label{cutoff_radius_numerical_Sun}
r_{\chi,\odot} \approx 0.01\,R_{\odot}\,\sqrt{\frac{\rm 100\,GeV}{m_\chi}},
\end{equation}
where $R_{\odot}$ is the solar radius, in agreement with~\cite{baratella2014}. For the Earth, $T_{\oplus} = 5700\,$K and $\bar{\rho}_{\oplus} = 1.2\times 10^4\,$kg/m$^3$, yielding
\begin{equation} \label{cutoff_radius_numerical_Earth}
r_{\chi,\oplus} \approx 0.1\,R_{\oplus}\,\sqrt{\frac{\rm 100\,GeV}{m_\chi}},
\end{equation}
where $R_{\oplus}$ is the Earth radius. Performing the integration in Eq.~\eqref{def_CA} with the potential in Eq.~\eqref{def_grav_potential} gives $C_A = \sann/\Ve$, where the effective volume is
\begin{equation} \label{effective_volume}
\Ve = \sqrt{2}\pi\,r_\chi^3\,\frac{\left[{\rm erf}\left(\frac{R}{r_\chi}\right)-\frac{R}{r_\chi}\,e^{-R^2/r_\chi^2}\right]^2}{{\rm erf}\left(\frac{\sqrt{2}R}{r_\chi}\right)-\frac{\sqrt{2}R}{r_\chi}\,e^{-2R^2/r_\chi^2}},
\end{equation}
and where
\begin{equation}
{\rm erf}(\zeta) = \int_0^\zeta\,e^{-t^2}\,dt.
\end{equation}

\section{Review of the WIMP capture rate} \label{Review of the WIMP capture rate}

We review the derivation of the capture rate $C$, following the seminal work in Refs.~\cite{press1985, freese1986, krauss1986, gould1987b}. In the derivation, we include the dependence of the scattering cross section on the recoil energy as in Refs~\cite{nussinov2009, catena2015}. The differential number of WIMPs with velocity within $\bu$ and $\bu + d^3\bu$ and in the volume element $d^3\bx$ is given by
\begin{equation}
dN_\chi = \Psi(t,\bx,\bu)d^3\bx\,d^3\bu,
\end{equation}
where $\Psi = \Psi(t, \bx, \bu)$ is the DM phase space distribution far away from compact objects, following the Liouville theorem
\begin{equation} \label{liouville}
\frac{d\Psi}{dt} = \frac{\partial \Psi}{\partial t} + \bu \cdot \n \Psi - \n\,\phi\cdot \frac{\partial \Psi}{\partial \bu} = 0.
\end{equation}
The number density is given in terms of the phase space distribution as
\begin{equation}
n_\chi = \int \Psi\,d^3\bu.
\end{equation}
We assume that the function $\Psi$ depends on $u = |\bu|$ and $\bx$ only, and introduce the velocity distribution $f(u)du = 4\pi\,\Psi\,u^2\,du$. Different forms of $f(u)$ have been discussed in Sec.~\ref{Velocity distribution}.
 
The inward differential WIMP flux across a shell of radius $R$ coming from a direction at an angle $\theta$ with respect to the radial direction is~\cite{press1985}
\begin{equation}
dF = \frac{1}{4}\,f(u)\,u\,du\,d(\cos^2\theta),
\end{equation}
from which the differential accretion rate is
\begin{equation} \begin{split}
d\mF &= 4\pi\,R^2\,dF = \pi\,R^2\,f(u)\,u\,du\,d(\cos^2\theta)
\\ &= \frac{\pi}{m_\chi^2} \, \frac{f(u)}{u}\,du\,dJ^2.
\end{split} \end{equation}
In the last expression, we have used the angular momentum of the particle $J = m_\chi\,R\,u\,\sin\theta$ as the integration variable in place of $\theta$. The velocity $w$ near the shell is given by the conservation of energy as
\begin{equation} \label{energy_cons}
w^2 = u^2 + \ve^2,
\end{equation}
where $\ve$ is the escape velocity at radius $r$. Following Ref.~\cite{gould1987b}, we define the rate $\Omega_w$ per unit time at which a WIMP with velocity $w$ scatters to a velocity less than $\ve$ in a shell at radius $r$ with width $dr$. The time spent within the shell is found by imposing energy conservation, and reads
\begin{equation}
dt = \frac{2dr}{w\,\sqrt{1-\left(\frac{J}{m_\chi\,r\,w}\right)^2}}\,\Theta\left(m_\chi\,r\,w - J\right).
\end{equation}
The specific capture rate, which is the number of WIMPs captured per unit time and unit volume, is
\begin{equation} \label{specificcapturerate} \begin{split}
\frac{dC}{dV} &= \frac{1}{4\pi\,r^2\,dr}\int_{J=0}^{J=+\infty} \Omega^s_w\,d\mF\,dt
\\ &= \int_0^{+\infty}\, \Omega^s_w\,w\, \frac{f(u)}{u}\,du,
\end{split} \end{equation}
where $s$ accounts for either SI or SD WIMP-nucleon scattering.

The rate $\Omega^s_w$ is the product of the probability $\Pi_w$ that a WIMP after the scattering has a velocity smaller than $\ve$ and the rate for scattering off the element $i$ given by
\begin{equation} \label{scattering_rate}
\frac{\Omega^s_w}{\Pi_w} = \int\, n_{i}\,w\,\frac{d\sigma^s_i}{dE_R}\,dE_R.
\end{equation}
In the last expression, $E_R$ is the nucleon recoil energy that the WIMP loses in the collision with a nucleus of species $i$ and number density $n_i$ in the body, and $\sigma^s_i$ is the WIMP-nucleon cross section for $s$ being either SI or SD. To find $\Pi_w$, we consider a WIMP with velocity $w$ and energy $E_w = m_\chi\,w^2/2$ scattering off a nucleus of mass $m_i$. The WIMP energy loss is
\begin{equation}
0 \leq \left|\frac{\Delta E_w}{E_w}\right| \leq \frac{4\mu_i^2}{m_\chi\,m_i},
\end{equation}
where $\mu_i = m_\chi\,m_i / (m_\chi + m_i)$ is the reduced mass, $\Delta E_w = E_w - E_w' = E_R$, with $E_R$ the recoil energy of the nucleus, and the upper bound is given by energy-momentum conservation. In order for the particle to be bound, the energy loss must fall in the range
\begin{equation} \label{prob_energyloss}
\frac{u^2}{w^2} \leq \left|\frac{\Delta E_w}{E_w}\right| \leq \frac{4\mu_i^2}{m_\chi\,m_i}.
\end{equation}
The probability that the energy loss falls in the range in Eq.~\eqref{prob_energyloss} is then
\begin{equation}\label{prob}
\Pi_w = \frac{m_\chi\,m_i}{4\mu_i^2}\,\left(\frac{4\mu_i^2}{m_\chi\,m_i} -\frac{u^2}{w^2}\right) \,\Theta\left(\frac{4\mu_i^2}{m_\chi\,m_i} -\frac{u^2}{w^2}\right),
\end{equation}
where the $\Theta$ function has been inserted to assure that $|\Delta E_w/E_w|$ is positive. The condition inside the $\Theta$ function converts into an upper limit for the velocity at infinity,
\begin{equation}
u \leq u_{\rm max} \equiv \ve\,\sqrt{\frac{4\,m_i\,m_\chi}{(m_i-m_\chi)^2}},
\end{equation}
so that Eq.~\eqref{prob} can be written as
\begin{equation}\label{prob1}
\Pi_w = \frac{\ve^2}{w^2}\,\left[1-\left(\frac{u}{u_{\rm max}}\right)^2\right] \,\Theta\left(u_{\rm max} - u\right).
\end{equation}
Substituting Eqs.~\eqref{scattering_rate} and~\eqref{prob1} into Eq.~\eqref{specificcapturerate}, and integrating over the volume of the body, gives the capture rate
\begin{equation} \label{capturerate0} \begin{split}
C &= \sum_i\,4\pi\,\int_0^R dr\,r^2\,\ve^2(r)\,n_i \times
\\ &\times \int_0^{u_{\rm max}}\,du\,\frac{f(u)}{u}\,\left[1-\left(\frac{u}{u_{\rm max}}\right)^2\right]\,\int_{\Em}^{\EM}dE_R\,\frac{d\sigma^s_i}{dE_R}.
\end{split} \end{equation}
Here, the limits of integration over the differential recoil energy $dE_R$ are
\begin{equation} \label{limits_integration}
\Em = \frac{1}{2}\,m_\chi\,u^2,\quad\hbox{and}\quad \EM = \frac{2\mu_i^2}{m_i}\,\left[u^2+\ve^2(r)\right].
\end{equation}
We replace the number density profile $n_i(r)$ of the element $i$ in the capturing body with the mass fraction $x_i$ via
\begin{equation}
x_i = \frac{1}{M}\,\int\,n_i\,m_i\,dV,
\end{equation}
where $M$ is the mass of the body. The knowledge of the distribution of the elements inside the capturing body is crucial in correctly determining the capture rate, as expressed in the integral over the volume of the capturing body in Eq.~\eqref{capturerate0}. In fact, the radial dependence of the integrand comes from the distribution of the material in the Sun and in the Earth and from the dependency of the escape velocity $\ve = \ve(r)$ on the distance from the core $r$, which can be approximated in terms of the mass enclosed in the radius $r$, $M(r)$, as~\cite{gould1992},
\begin{equation} \label{escape_velocity_radius}
\ve^2(r) = \ve^2(0) - \frac{M(r)}{M}\,\left(\ve^2(0)-\ve^2(R)\right).
\end{equation}
Here, instead of performing the integration over the radius $r$, we use the approximation outlined in Ref.~\cite{jungman1996}, where the authors introduce a new quantity $\phi_i$ which describes the gravitational potential of element $i$ in the Sun or the Earth relative to the surface,
\begin{equation} \label{definition_phi}
\phi_i = \frac{\int\,\ve^2(r)\,\rho_i\,dV}{\ve^2(R)\,x_i\,M}.
\end{equation}
Eq.~\eqref{definition_phi} neglects the radial dependence of the bounds of integration over the recoil energy in Eq. \eqref{limits_integration}, which are computed at $r=R$. With this approximation, the capture rate in Eq.~\eqref{capturerate0} is
\begin{equation} \label{rateC} \begin{split}
C &= \frac{\ve^2(R)}{v_\sigma}\,\frac{\rho_\chi}{m_\chi}\,M\,\sum_i\,\frac{x_i}{m_i}\,\phi_i \times
\\ &\times \int_0^{\xi_{\rm max}}\,\frac{d\xi}{\xi}\,\tf(\xi)\,\left[1-\left(\frac{\xi}{\xi_{\rm max}}\right)^2\right]\,\int_{\Em}^{\EM}\,\frac{d\sigma^s_i}{dE_R}\,dE_R,
\end{split} \end{equation}
where
\begin{equation}
\xi_{\rm max} = \sqrt{\frac{3}{2}}\,\frac{u_{\rm max}}{v_\sigma},
\end{equation}
is the maximum value of $\xi$ for which the quantity in square brackets is positive, and all quantities that depend on the radius are computed at $r=R$, and $\rho_\chi = m_\chi\,n_\chi$ is the WIMP energy density.

The differential cross section $d\sigma^s_i/dE_R$ is reviewed in Appendix~\ref{Differential cross sections} below. Here, we anticipate the relevant result in Eq.~\eqref{cross_section_omega} which, once inserted into Eq.~\eqref{rateC}, allows us to express the capture rate as
\begin{equation} \label{capture_rate_general}
C = K^s(m_\chi)\,\sigma^s_p\,\rho_\chi,
\end{equation}
where $\sigma^s_p$ is the WIMP-proton scattering cross section at zero momentum for either SI or SD and
\begin{equation} \label{def_kappa} \begin{split}
K^s(m_\chi) &= \frac{M}{2\,m_\chi\,\mu_p^2\,v_\sigma}\,\sum_i\,\omega^s_i\,x_i\,\phi_i\times
\\ &\times \int_0^{\xi_{\rm max}}\,\frac{d\xi}{\xi}\,\tf(\xi)\,\int_{\Em}^{\EM}\,F_s^2(E_R)\,dE_R.
\end{split} \end{equation}
Here $\omega^s_i$, defined in Eq.~\eqref{def_omega}, describes the enhancement due to the number $A_i$ of nucleons in the nuclei $i$ for SI, and due to the total nucleon spin $J_i$ for SD, $\mu_p$ is the WIMP-proton reduced mass, and $F_s(E_R)$ is the form factor.

\section{Velocity distribution} \label{Velocity distribution}

The velocity distribution $f(u)$ is a solution to the stationary Liouville equation, as we review in Appendix~\ref{Review of the WIMP capture rate}, see Eq.~\eqref{liouville}. In the galactic rest frame, the velocity follows a Maxwell-Boltzmann distribution~\cite{gould1987b} according to the standard DM halo model~\cite{drukier1986, freese1988},
\begin{equation} \label{phase_density}
f(u) = 4\pi\,n_\chi\,\left(\frac{3}{2\pi\,v_\sigma^2}\right)^{3/2}\,u^2\,e^{-\xi^2},
\end{equation}
where $\xi = \sqrt{3/2}\,u/v_\sigma$ and $v_\sigma = 270\,$km/s is a velocity dispersion\footnote{Although there has been concern
that the velocity distribution of the DM might deviate significantly from Maxwellian $f(u)$ in Eq.~\eqref{phase_density}, Refs. ~\cite{kelso2016, bozorgnia, sloane2016} showed that results obtained for DM with a Maxwellian profile are consistent to those obtained when baryons are included in DM simulations, though there is as yet possible disagreement for the high velocity tail.}. Boosting to the Sun's rest frame with velocity $v_{\odot}$ relative to the galactic rest frame, Eq.~\eqref{phase_density} becomes
\begin{equation} \label{phase_density1}
f(\xi) = \frac{n_\chi}{v_\sigma}\,\tf(\xi) = \frac{n_\chi}{v_\sigma}\,\sqrt{\frac{24}{\pi}}\,\xi^2\,\left(\frac{\sinh\,2\xi\eta}{2\xi\eta}\right)\,e^{-\xi^2-\eta^2},
\end{equation}
where
\begin{equation}
\eta = \sqrt{\frac{3}{2}}\,\frac{v_{\odot}}{v_\sigma},
\end{equation}
and $\tf(\xi)$ is dimensionless. Since the capture time $\tann$ is much greater than 1\,yr, we can average over the motion of the Earth around the Sun and thus use the velocity distribution Eq. \eqref{phase_density1} for capture in both the Sun and the Earth. It has been shown~\cite{sivertsson2010, sivertsson2012} that the changes of the velocity distribution at the Earth from WIMP capture in neighboring massive bodies such as other planets or the Sun (cf.~\cite{peter2009a, peter2009b}) is negligible.

\section{Differential cross sections} \label{Differential cross sections}

\subsection{Capture rate for spin-independent interaction} \label{Capture for spin-independent interaction}

The SI cross section of a WIMP off the nucleus species $i$ (with $A_i$ nucleons of which $Z_i$ are protons) at zero momentum transfer is~\cite{jungman1996}
\begin{equation} \label{def_sigmai} \begin{split}
\sigma_i^{\rm SI}(0) &= \frac{4}{\pi}\,\mu_i^2\,\left[Z_i \,f_p + (A_i-Z_i)\,f_n\right]^2
\\ &\approx A_i^2\,\left(\frac{\mu_i}{\mu_p}\right)^2\,\SI.
\end{split} \end{equation}
Here, $\SI$ is the WIMP-proton cross section at zero momentum transfer, which is the quantity bound by direct-detection experiments~\cite{agnese2014, akerib2014, amole2016}, and $f_p$ ($f_n$) is a model-dependent quantity parametrizing the WIMP-proton (-neutron) matrix element. For the last approximation we assume $f_n = f_p$. The SI differential cross section is obtained using Eq.~\eqref{def_sigmai} as
\begin{equation} \label{def_SI}
\frac{d\sigma^{\rm SI}_i}{dE_R} = \frac{\sigma^{\rm SI}_i(0)}{\EM - \Em}\,\FSI^2(E_R) = \frac{A_i^2\,m_i\,\SI}{2\,\mu_p^2\,\ve^2}\,\frac{\FSI^2(E_R)}{1-\left(\frac{\xi}{\xi_{\rm max}}\right)^2},
\end{equation}
where for the SI interaction we use the Helm\footnote{More refined nuclear form factors for each nuclear interaction have recently been computed in Ref.~\cite{catena2015}. Comparing the form factors computed in various model usually yields negligible changes at small WIMP masses $m_\chi \lesssim 10\,$GeV, while for larger WIMP masses one finds  ${\cal O} (10 \%)$ differences in the scattering rates~\cite{Joakim}.} nuclear form factor~\cite{helm1956}
\begin{equation} \label{form_factor_SI}
\FSI(E_R) = e^{-E_R/E_i},
\end{equation}
with energy cutoff and nuclear radius given by~\cite{gould1987b}
\begin{equation} \label{energy_cutoff_nuclear_radius}
E_i = \frac{3\hbar^2}{2\,m_i\,R_i^2},\quad\hbox{and}\quad R_i = \left[0.91\,\left(\frac{m_i}{\rm GeV}\right)^{1/3} + 0.3\right]\,{\rm fm}.
\end{equation}

\subsection{Capture rate for spin-dependent interaction}

WIMPs can couple to the nucleus via spin-spin interaction, giving rise to spin-dependent (SD) scattering. We model the SD cross section as~\cite{gondolo1996, jungman1996, tovey2000, savage2004}
\begin{equation} \label{SD_diff_crosssection} \begin{split}
\frac{d\sigma^{\rm SD}_i}{dE_R} &= \frac{\sigma^{\rm SD}_i(0)}{\EM - \Em}\,\FSD^2(E_R)
\\ &= \frac{16m_i\,G_F^2}{\pi\,\ve^2}\,\frac{J_i+1}{J_i}\times
\\ &\times \left(a_p\,\Sp + a_n\,\Sn\right)^2\,\frac{\FSD^2(E_R)}{1-\left(\frac{\xi}{\xi_{\rm max}}\right)^2},
\end{split} \end{equation}
where $a_p$ ($a_n$) is a dimensionless model-dependent quantity which takes the role of $f_p$ ($f_n$) in Eq.~\eqref{def_sigmai}, defined in terms of the WIMP-proton (-neutron) cross section $\SDp$ ($\SDn$) at zero momentum transfer,
\begin{equation}
\SDp = \frac{24}{\pi}\,G_F^2\,\mu_p^2\,a_p^2, \qquad \SDn = \frac{24}{\pi}\,G_F^2\,\mu_n^2\,a_n^2.
\end{equation}
$\Sp$ and $\Sn$ are respectively the expectation values of the proton and neutron spins within the nucleus $i$ with total nuclear spin $J_i$, $\FSD(E_R)$ is the form factor as a function of the recoil energy $E_R$~\cite{lewin1996}, and $G_F = 1.17 \times 10^{-5}\,$GeV$^{-2}$ is a constant. The spin expectation values are computed using detailed nuclear physics models. Here, we use the zero-momentum spin structure obtained from the extended odd group model~\cite{engel1989, engel1993} as tabulated in Ref.~\cite{bednyakov2004} where available. For isotopes not listed in Ref.~\cite{bednyakov2004}, we use the results from the odd group model~\cite{engel1989, engel1993}. Other nuclear shell models like the independent single particle shell model~\cite{ellis1988, ellis1991} and the interacting boson-fermion model~\cite{iachello1991} exist, with different techniques often yielding different results. See Refs.~\cite{jungman1996, bednyakov2004} for a review of the effects of these models on WIMP direct detection experiments.

Assuming that the SD cross sections of WIMPs off neutrons and protons are equal, it is convenient to rewrite Eq.~\eqref{SD_diff_crosssection} as 
\begin{equation} \label{SD_diff_crosssection1}
\frac{d\sigma^{\rm SD}_i}{dE_R} =  \frac{\lambda_i^2\,m_i\,\SDp}{2\,\mu_p^2\,\ve^2}\,\frac{\FSD^2(E_R)}{1-\left(\frac{\xi}{\xi_{\rm max}}\right)^2},
\end{equation}
where the model dependency is absorbed into
\begin{equation} \label{lambda_i_def} \begin{split}
\lambda_i^2 &\equiv \frac{4}{3}\frac{J_i+1}{J_i}\,\left(\Sp + {\rm sign}(a_p, a_n)\,\Sn\right)^2 
\\ &\approx \frac{4}{3}\,\frac{J_i+1}{J_i}\,\left(\Sp + {\rm sign}(a_p, a_n)\,\Sn\right)^2.
\end{split} \end{equation}
There are several important differences between the form of the SI and SD cross sections that greatly affect the capture rate:
\begin{itemize}
\item The enhancement $A_i^2$ appearing in Eq.~\eqref{def_sigmai} for SI interaction is replaced by $(J_i+1)/J_i$ in Eq.~\eqref{SD_diff_crosssection}, so heavier nuclei do not enhance SD capture as for capture via SI scattering.

\item Not all nuclei but only those with $J_i \neq 0$ have non-vanishing SD interactions.

\item Experimentally the proton-WIMP and neutron-WIMP SD cross sections are not as tightly constrained as $\SI$, see Sec.~\ref{Are the Sun or the Earth in equilibrium?} below.

\item The dependence of the SD form factor on $E_R$ differs from the SI approximation in Eq.~\eqref{form_factor_SI}. Here, we use the expression given in Ref.~\cite{lewin1996, klapdor2005}, obtained within the ``thin shell'' approximation and valid for all nuclei,
\begin{equation}
\FSD^2(E_R) = \begin{cases}
j_0^2(x) &\hbox{for $x\leq 2.55$ or $x\geq 4.5$},\\
0.047 & \hbox{for $2.55<x<4.55$},
\end{cases}
\end{equation}
where $j_n(x)$ is a spherical Bessel function of the first kind and $x \equiv R_1\,E_R$, with the effective nuclear radius~\cite{lewin1996, helm1956}
\begin{equation}
R_1 = \sqrt{(1.23 A^{1/3}-0.6)^2 + 2.177}\,{\rm fm}.
\end{equation}
\end{itemize}

\subsection{General expression for the differential scattering rate}

The expressions for the SI and SD scattering in Eqs.~\eqref{def_SI} and~\eqref{SD_diff_crosssection} can be combined in the general form~\cite{nussinov2009}
\begin{equation} \label{cross_section_omega}
\frac{d\sigma^s_i}{dE_R} = \frac{\sigma^s_i(0)}{\EM - \Em}\,F_s^2(E_R) = \frac{\omega^s_i\,m_i\,\sigma^s_p}{2\,\mu_p^2\,\ve^2}\,\frac{F_s^2(E_R)}{1-\left(\frac{\xi}{\xi_{\rm max}}\right)^2},
\end{equation}
where the label $s$ stands for either SI or SD scattering, $\sigma^s_i(0)$ is the WIMP-nucleon cross section at zero momentum transfer for the nucleus $i$, $F_s(E_R)$ is a nuclear form factor accounting for the finite size of the nucleus,
\begin{equation} \label{def_omega}
\omega^s_i = \begin{cases}
A_i^2 & \hbox{for SI},\\
\lambda_i^2 &\hbox{for SD},
\end{cases}
\end{equation}
and where $\lambda_i$ has been defined in Eq.~\eqref{lambda_i_def}. The denominator $\EM - \Em$ in Eq.~\eqref{cross_section_omega} is chosen so that, when $F_s(E_R) = 1$,
\begin{equation}
\int_{\Em}^{\EM}\,\frac{d\sigma^s_i}{dE_R}\,dE_R = \sigma^s_i(0).
\end{equation}
For this reason, Eq.~\eqref{cross_section_omega} differs from the corresponding expression in the direct detection literature, where $\Em = 0$ and hence
\begin{equation}
\left(\frac{d\sigma^s_i}{dE_R}\right)_{\rm DD} = \frac{m_i\,\sigma^s_i(0)}{2\,\mu_i^2\,w^2}\,F_s^2(E_R).
\end{equation}

\section{Computation of the WIMP relic density} \label{appendix_relic}

\renewcommand{\thefigure}{G\arabic{figure}}

\setcounter{figure}{0}

In this Appendix, we compute the thermally-averaged annihilation cross-section in the Early Universe $\left\langle\sigma v\right\rangle_{\rm ann, EU}(m_\chi; \fc)$ assuming standard freeze-out production necessary for WIMPs to make up a fraction $\fc = \Omega_\chi/\Omega_{\rm DM}$ (cf Eq.~\eqref{subdominant_energy_density}) of the total DM for a given WIMP mass $m_\chi$. As long as co-annihilation plays no important role, $\left\langle\sigma v\right\rangle_{\rm ann, EU}$ differs from $\sann$ in the Sun or the Earth used to compute the annihilation rate only by the temperature at freeze-out being different than today in the Sun or the Earth. For this work, we assume $s$-wave annihilation and neglect co-annihilation, yielding $\left\langle\sigma v\right\rangle_{\rm ann, EU} = \sann$.

We follow the computation of \cite{SDB12} for the WIMP relic density as a function of the velocity averaged annihilation cross-section $\sann$ and the WIMP mass $m_\chi$, assuming the WIMPs to be in thermal equilibrium before freeze-out. Under those assumptions, the relic density can be written as \cite{SDB12}
\begin{equation} \label{OmFreeze} \Omega_\chi h^2 = \frac{9.92 \times 10^8}{\sann} \left(\frac{x_*}{g_*^{1/2}}\right)\left(\frac{\left(\Gamma_A/H\right)_*}{1+\alpha_* \left(\Gamma_A/H\right)_*}\right) \ , \end{equation}
where $\Omega_\chi \equiv \rho_\chi/\rho_c$ with $\rho_c = 3H_0^2/8 \pi G$ is the WIMP energy density $\rho_\chi$ in terms of the critical density $\rho_c$, and $h = H_0/(100\,{\rm km\,s}^{-1}{\rm Mpc}^{-1})$ is the reduced Hubble constant. We use $x \equiv m/T$ where $T$ is the temperature as a proxy for time. $g = g(T)$ measures the relativistic degrees of freedom. $\Gamma_A = n_\chi\,\sann$ is the annihilation rate of WIMPs where $n_\chi = \rho_\chi/m_\chi$ is the number density. We parametrize the deviation from thermal equilibrium by $\Delta$ via $n \equiv \left(1+\Delta\right)n_{\rm eq}$. Starred quantities are calculated when
\begin{equation} \frac{\Delta(x_*)\left(2+\Delta(x_*)\right)}{\left(1+\Delta(x_*)\right)} = 0 \ ,\end{equation}
hence, shortly after departure from equilibrium when $\Delta_* \simeq 0.618$ or $n \simeq 1.618\,n_{\rm eq}$. The effect of changing of $g(T)$ is taken into account via the integral
\begin{equation} \alpha_* \equiv \int_{T_f}^{T_*} \frac{dT}{T_*} \sqrt{\frac{g}{g_*}}\left(1+\frac{1}{3}\frac{d(\ln g)}{d (\ln T)} \right), \end{equation}
where $T_f$ is the present temperature, but as \cite{SDB12} we use $T_f = T_*/100$ in our calculations since the largest contributions to the integral come from $T \sim T_*$. 

The relic density as a function of  $\sann$ and $m_\chi$ is computed by first solving
\begin{equation} \begin{split}
& x_* + \ln(x_* - 3/2) - 0.5 \ln x_* = 
\\ & = 20.5 + \ln(<\sigma v>_{\rm ann}/10^{-26}\,{\rm cm}^3\,{\rm s}^{-1}) + 
\\ & + \ln(m_\chi/{\rm GeV}) - 0.5 \ln g_*\end{split} \end{equation}
to obtain $x_*$. One then goes on to calculate $g_*$ and $\alpha_*$. The annihilation rate is given by
\begin{equation}
\left(\frac{\Gamma_A}{H}\right)_* \simeq \left(1+\Delta_*\right) \frac{x_* - 3/2 - \frac{d(\ln g)}{d(\ln T)}}{1+\frac{1}{3} \frac{d(\ln g)}{d(\ln T)}}.
\end{equation}

Fig. \ref{fig:relDens_1} shows $\sann$ as a function of $m_\chi$ required for WIMPs to make up a fraction $\fc$ of DM. For a given $f_\chi$, we recover the well-known behavior that $\sann$ is almost independent of the WIMP mass for $m_\chi \gtrsim 30\,$GeV. For smaller masses, $\sann$ becomes dependent on $m_\chi$: The freeze-out temperature is roughly given by $T_f \approx m_\chi/20$. Thus, for smaller masses the number of relativistic degrees of freedom at decoupling $g_*$ changes, because $b$-quarks become non-relativistic at $T \sim m_b =  4.2\,$GeV and $c$-quarks at $T \sim m_c = 1.3\,$GeV. For even lighter WIMPs, the QCD phase transition sets in at $T \approx \Lambda_{\rm QCD} \approx 220\,$MeV, reducing $g_*$ before the WIMPs decouple.

For a given WIMP mass, we na{\"i}vely expect scaling $\sann \propto 1/\fc$ from Eq. \eqref{OmFreeze}. The deviations from this scaling, which are of the order of 20\,\% for $\fc = 1\%$ when compared to $\fc = 100\,\%$, are again caused by the changing number of effective degrees of freedom at freeze-out, since particles with the same mass but larger $\sann$ freeze-out later and hence at smaller $g_*$.

\end{document}